\documentclass[11pt]{article}
\usepackage{amsthm,amsmath,amssymb,a4wide}
\usepackage{psfrag,graphics}

\newcommand{\yo}[1][]{y_{\bot,j}}
\newcommand{\dyo}[1][]{\dot{y}_{\bot,j}}

\newcommand{\vel}{v}

\newcommand{\sol}{Q}

\newcommand{\set}[1]{\mathrm{#1}}

\newcommand{\RE}{\mathop{\set{Re}}}

\newcommand{\ie}{{i.\,e.\/}, }

\newcommand{\eg}{{e.\,g.\/}, }
\newcommand{\diff}{\mathop{\mathrm{\mathstrut{d}}}\!}%Diff op.

\newcommand{\Hone}{\set{H}^1}

\newcommand{\Sob}[1]{\set{H}^{#1}}
\newcommand{\C}[1]{\set{C}^{#1}}

\newcommand{\Lp}[1]{\set{L}^{#1}}

\newtheorem{proposition}{Proposition}[section]

\newtheorem{theorem}{Theorem}
\newtheorem{lemma}[proposition]{Lemma}

\newcommand{\En}{\mathcal{E}}

\newcommand{\Ev}{\En_{\vel}}

\newcommand{\Tm}{\big ( \sqrt{-\Delta+m^2} - m \big ) }

\newcommand{\Tmu}{\big ( \sqrt{-\Delta + m^2} - m + \mu \big ) }
\newcommand{\T}{\sqrt{-\Delta}}
\newcommand{\Vconv}[1]{ \big ( \frac{1}{|x|} \ast |#1|^2 \big ) #1 }
\newcommand{\RR}{\mathbb{R}}
\newcommand{\Hhalf}{\set{H}^{1/2}}

\newcommand{\Lploc}[1]{\set{L}^{#1}_{\set{loc}}}

\newcommand{\inner}[2]{  \langle #1, #2  \rangle }
\newcommand{\innerb}[2]{ \big \langle #1, #2 \big \rangle }
\newcommand{\Nc}{N_{\mathrm{c}}}
\newcommand{\dd}{\set{d}}
\newcommand{\Vpot}[1]{ \int_{\RR^3} \big ( \frac{1}{|x|} \ast |#1|^2 \big ) |#1|^2 \diff x }

\newcommand{\nrmHhalf}[1]{ \| #1 \|_{\set{H}^{1/2}}}
\newcommand{\Nfun}{\mathcal{N}}

\renewcommand{\sol}{\varphi}
\renewcommand{\En}{\mathcal{E}}

\numberwithin{equation}{section}

\begin{document}

\title{{\bf Boson Stars as Solitary Waves}}

\author{J\"urg Fr\"ohlich \and B.~Lars G.~Jonsson \and Enno Lenzmann}
%\footnote{$^1$ Institute for Theoretical Physics, ETH Z\"urich--H\"onggerberg, CH-8093 Z\"urich, Switzerland}

%\address{$^2$ Division of Electromagnetic Theory, Alfv\'en Laboratory, Royal Institute of Technology, SE-100 44 Stockholm, Sweden}

%x\address{$^3$ Department of Mathematics, ETH Z\"urich--Zentrum, CH-8092 Z\"urich, Switzerland}
%\ead{lenzmann@math.ethz.ch}
\date{December 12, 2005}
\maketitle

\begin{abstract}
We study the nonlinear equation 
\[
i \partial_t \psi = \Tm \psi - ( |x|^{-1} \ast |\psi|^2 ) \psi \quad \mbox{on $\RR^3$},
\] 
which is known to describe the dynamics of pseudo-relativistic boson stars in the mean-field limit. For positive mass parameters, $m > 0$, we prove existence of travelling solitary waves, $\psi(t,x) = e^{i t \mu} \sol_{v}(x-vt)$, with speed $|v| < 1$, where $c=1$ corresponds to the speed of light in our units. Due to the lack of Lorentz covariance, such travelling solitary waves cannot be obtained by applying a Lorentz boost to a solitary wave at rest (with $v=0$). To overcome this difficulty, we introduce and study an appropriate variational problem that yields the functions $\sol_v \in \Hhalf(\RR^3)$ as minimizers, which we call boosted ground states. Our existence proof makes extensive use of concentration-compactness-type arguments. 

In addition to their existence, we prove orbital stability of travelling solitary waves $\psi(t,x) = e^{it \mu} \sol_v(x-vt)$ and pointwise exponential decay of $\sol_v(x)$ in $x$.  

%In a companion paper \cite{FJL2005II} the main result of which is outlined in the last section of this paper, we study the effective dynamics of travelling solitary waves in an external potential, over a finite time interval. 
\end{abstract}

%\submitto{\NL}

\section{Introduction}
\label{sec-introduction}

In this paper and its companion \cite{FJL2005II}, we study solitary wave solutions --- and solutions close to such --- of the pseudo-relativistic Hartree equation
\begin{equation} \label{eq-hartree}
i \partial_t \psi = \Tm \psi - \Vconv{\psi} \quad \mbox{on $\RR^3$.}
\end{equation} 
Here $\psi(t,x)$ is a complex-valued wave field, and the symbol $\ast$ stands for convolution on $\RR^3$. The operator $\sqrt{-\Delta + m^2}-m$, which is defined via its symbol $\sqrt{k^2+m^2}-m$ in Fourier space, is the kinetic energy operator of a relativistic particle of mass, $m \geq 0$, and the convolution kernel, $|x|^{-1}$, represents the Newtonian gravitational potential in appropriate physical units. 

As recently shown by Elgart and Schlein in \cite{Elgart+Schlein2005}, equation (\ref{eq-hartree}) arises as an effective dynamical description for an $N$-body quantum system of relativistic bosons with two-body interaction given by Newtonian gravity. Such a system is a model system for a pseudo-relativistic {\em boson star}. That is, we consider a regime, where effects of special relativity (accounted for by the operator $\sqrt{-\Delta +m^2}-m$) become important, but general relativistic effects can be neglected. The idea of a mathematical model of pseudo-relativistic boson stars dates back to the works of Lieb and Thirring \cite{Lieb+Thirring1984} and of Lieb and Yau \cite{Lieb+Yau1987}, where the corresponding $N$-body Hamiltonian and its relation to the Hartree energy functional $\mathcal{H}(\psi) = 2\mathcal{E}(\psi)$ are discussed, with $\mathcal{E}(\psi)$ defined in (\ref{eq-hartree-E}), below.

Let us briefly recap the state of affairs concerning equation (\ref{eq-hartree}) itself. With help of the conserved quantities of charge, $\Nfun(\psi)$, and energy, $\mathcal{E}(\psi)$, given by 
\begin{equation} \label{def-Nfun}
\Nfun(\psi)  = \int_{\RR^3} | \psi |^2 \diff x, 
\end{equation}
\begin{equation} \label{eq-hartree-E}
\mathcal{E}(\psi)  = \frac{1}{2} \int_{\RR^3} \overline{\psi} \Tm \psi \diff x - \frac{1}{4} \Vpot{\psi}, 
\end{equation}   
results derived so far can be summarized as follows (see also Fig.~\ref{fig-1} below).  
\begin{itemize}
\item {\bf Well-Posedness:}  For any initial datum $\psi_0 \in \Hhalf(\RR^3)$, there exists a unique solution 
\begin{equation}
\psi \in \C{0} \big ([0,T); \Hhalf(\RR^3) \big ) \cap \C{1} \big ( [0,T); \Sob{-1/2}(\RR^3) \big ), 
\end{equation}
for some $T > 0$, where $\Sob{s}(\RR^3)$ denotes the inhomogeneous Sobolev space of order $s$. Moreover, we have global-in-time existences (\ie  $T = \infty$) whenever the initial datum satisfies the condition
\begin{equation} \label{ineq-Nc}
\Nfun(\psi_0) < \Nc,
\end{equation}
where $\Nc > 4/\pi$ is some universal constant; see \cite{Lenzmann2005LWP} for a detailed study of the Cauchy problem for (\ref{eq-hartree}) with initial data in $\Sob{s}(\RR^3)$, $s \geq 1/2$. 
\item {\bf Solitary Waves:} Due to the focusing nature of the nonlinearity in (\ref{eq-hartree}), there exist solitary wave solutions, which we refer to as {\em solitary waves}, given by  
\begin{equation} \label{eq-sol-intro}
\psi(t,x) = e^{i t \mu } \sol(x),
\end{equation}
where $\sol \in \Hhalf(\RR^3)$ is defined as a minimizer of $\mathcal{E}(\psi)$ subject to $\Nfun(\psi) = N$ fixed. Any such minimizer, $\sol(x)$, is called a {\em ground state} and it has to satisfy the corresponding Euler-Lagrange equation
\begin{equation} \label{eq-sol-EL}
\Tm \sol - \big ( \frac{1}{|x|} \ast |\sol|^2 ) \sol = - \mu \sol,
\end{equation}
for some $\mu \in \RR$. An existence proof of ground states, for $0 < \Nfun(\sol) < \Nc$ and $m >0$, can be found in \cite{Lieb+Yau1987}. The method used there is based on rearrangement inequalities that allow one to restrict ones attention to radial functions, which simplifies the variational calculus. But in order to extend this existence result to so-called {\em boosted ground states}, \ie $x$ in (\ref{eq-sol-intro}) is replaced by $x-vt$ and equation (\ref{eq-sol-EL}) acquires the additional term, $i ( v \cdot \nabla ) \sol$, we have to employ concentration-compactness-type methods; see Theorem \ref{th-existence} and its proof, below. 

\item {\bf Blow-Up:} Any spherically symmetric initial datum, $\psi_0 \in \C{\infty}_{\mathrm{c}}(\RR^3)$, with 
\begin{equation} \label{ineq-blowup}
\mathcal{E}(\psi_0) < -\frac{1}{2}m \Nfun(\psi_0)
\end{equation}
leads to blow-up of $\psi(t)$ in a finite time, \ie we have that $\lim_{t \nearrow T} \nrmHhalf{\psi(t)} = \infty$ holds, for some $T < \infty$. We remark that (\ref{ineq-blowup}) implies that the smallness condition (\ref{ineq-Nc}) cannot hold. See \cite{Froehlich+Lenzmann2005} for a proof of this blow-up result.\footnote{In \cite{Froehlich+Lenzmann2005} the energy functional, $\mathcal{E}(\psi)$, is shifted by $+\frac{1}{2}m\mathcal{N}(\psi)$. Thus, condition (\ref{ineq-blowup}) reads $\mathcal{E}(\psi_0) < 0$ in \cite{Froehlich+Lenzmann2005}.}  In physical terms, finite-time blow-up of $\psi(t)$ is indicative of {\em ``gravitational collapse''} of a boson star modelled by (\ref{eq-hartree}); the constant $\Nc$ appearing in (\ref{ineq-Nc}) may then be regarded as a {\em ``Chandrasekhar limit mass''}. 
\end{itemize}

\begin{figure}[h]
\label{fig-1}
\begin{center}
 \psfrag{E1}{$E$}
 \psfrag{E2}{$E=-\frac{1}{2} mN$}
 \psfrag{N1}{$N$}
 \psfrag{N2}{$N=\Nc$}
 \psfrag{0}{0}
 \includegraphics{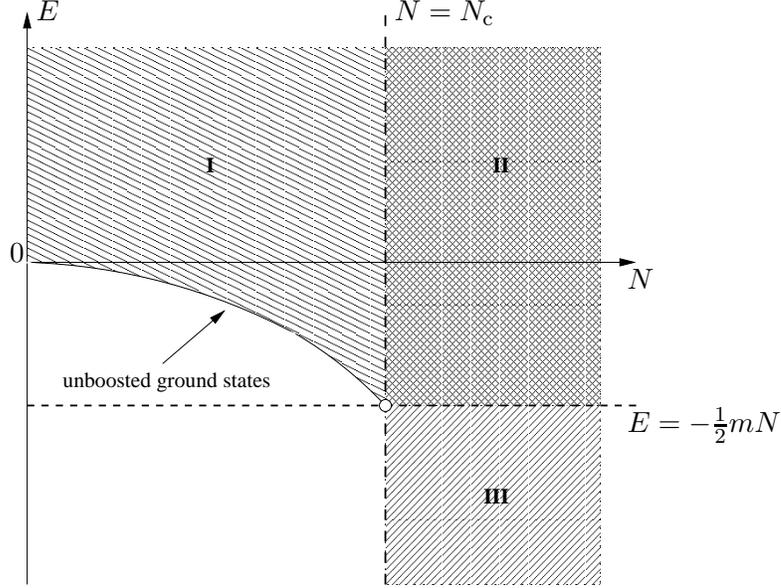}
\end{center}
\caption{Qualitative diagram for the boson star equation (\ref{eq-hartree}) with positive mass parameter $m > 0$. Here $N = \mathcal{N}(\psi_0)$ and $E = \mathcal{E}(\psi_0)$ denote charge and energy for the initial condition $\psi_0 \in \Hhalf(\RR^3)$. In region {\bf I}, all solutions are global in time and the (unboosted) ground states are minimizers of $\mathcal{E}(\psi)$ subject to fixed $\Nfun(\psi_0)=N$ with $0 < N < \Nc$. If $N$ exceeds $\Nc$, the energy $E$ can attain values below $-\frac{1}{2} mN$. As shown in \cite{Froehlich+Lenzmann2005} for spherically symmetric $\psi_0 \in \C{\infty}_{\mathrm{c}}(\RR^3)$ that belong to region {\bf III}, we have in fact blow-up of $\psi(t)$ within a finite time. Finally, the qualitative behavior of solutions with initial conditions in region {\bf II} appear to be of indefinite nature.}
\end{figure}

We now come to the main issues of the present paper which focuses on existence and properties of travelling solitary waves for (\ref{eq-hartree}). More precisely, we consider solutions of the form
\begin{equation} \label{eq-ansatz}
\psi(t,x) = e^{it \mu} \sol_v(x-vt)
\end{equation}
with some $\mu \in \RR$ and travelling velocity, $v \in \RR^3$, such that $|v| < 1$ holds (\ie below the speed of light in our units). We point out that, since equation (\ref{eq-hartree}) is not Lorentz covariant, solutions such as (\ref{eq-ansatz}) cannot be directly obtained from solitary waves at rest (\ie we set $v=0$) and then applying a Lorentz boost. To circumvent this difficulty, we plug the ansatz (\ref{eq-ansatz}) into (\ref{eq-hartree}). This yields
\begin{equation} \label{eq-boost-intro}
\Tm \sol_v + i (v \cdot \nabla) \sol_v - \big ( \frac{1}{|x|} \ast |\sol_v|^2 \big ) \sol_v = - \mu \sol_v ,
\end{equation}
which is an Euler-Lagrange equation for the following minimization problem
\begin{equation} \label{pr-Ev-intro}
\Ev(\psi) := \mathcal{E}(\psi) + \frac{i}{2} \int_{\RR^3} \overline{\psi} (v \cdot \nabla) \psi \diff x = \min ! \quad \mbox{subject to} \quad \Nfun(\psi) = N.
\end{equation}
We refer to such minimizers, $\sol_v \in \Hhalf(\RR^3)$, as {\em boosted ground states} throughout this paper. Indeed, we will prove existence of boosted ground states when $|v| < 1$ and $0 < N < \Nc(v)$ holds, as well as non-existence when $N \geq \Nc(v)$; see Theorem \ref{th-existence}, below. Our existence proof rests on concentration-compactness arguments which for our problem need some technical modifications, due to the pseudo-differential operator $\sqrt{-\Delta + m^2}$.  

Apart from existence of boosted ground states, we are also concerned with properties such as {\em ``orbital stability''} and exponential decay of $\sol_v(x)$ in $x$; see Theorems \ref{th-stability} and \ref{th-reg-decay}, below. We remark that both properties rely crucially on the positivity of the mass parameter, \ie we have $m > 0$ in (\ref{eq-hartree}). By contrast, it is shown, for instance, in \cite{Froehlich+Lenzmann2005} that (resting) solitary waves become unstable when $m=0$, due to nearby initial data leading to blow-up solutions.

In a companion paper \cite{FJL2005II}, we will explore the effective dynamics of (slowly) travelling solitary waves in an external potential; see also Sect.~\ref{sec-outlook} for a short summary of these result.

The plan of this paper is as follows.
\begin{itemize}
\item In Sect.~\ref{sec-existence}, we set-up the variational calculus for problem (\ref{pr-Ev-intro}) and we prove existence of boosted ground states, $\sol_v  \in \Hhalf(\RR^3)$, for $0 < \Nfun(\sol_v) < \Nc(v)$ and $|v| < 1$, as well as their nonexistence if $\mathcal{N}(\sol_v) \geq \Nc(v)$; see Theorem \ref{th-existence}, below.
\item Sect.~\ref{sec-stability} addresses ``orbital stability'' of travelling solitary waves $\psi(t,x) = e^{it\mu} \sol_v(x-vt)$; see Theorem \ref{th-stability}, below.
\item In Sect.~\ref{sec-properties}, we derive pointwise exponential decay and regularity of boosted ground states; see Theorem \ref{th-reg-decay}, below.
\item In Sect.~\ref{sec-outlook}, we sketch the main result of \cite{FJL2005II} describing the effective dynamics of travelling solitary waves in an external potential. 
\item In App.~A--C, we collect and prove several technical statements which we refer to throughout this text. 
\end{itemize}

\subsection*{Notation}

Lebesgue spaces of complex-valued functions on $\RR^3$ will be denoted by $\Lp{p}(\RR^3)$, with norm $\| \cdot \|_p$ and $1 \leq p \leq \infty$. Sobolev spaces, $\Sob{s}(\RR^3)$, of fractional order $s \in \RR$ are defined by
\begin{equation}
\Sob{s}(\RR^3) := \big \{ f \in \Lp{2}(\RR^3) : \| f \|^2_{\set{H}^s} := \int_{\RR^3} |\widehat{f}(k)|^2 (1 + |k|^2)^s \diff k < \infty \big \},
\end{equation}
where 
\begin{equation}
\widehat{f}(k) = \frac{1}{(2 \pi)^{3/2}} \int_{\RR^3} f(x) e^{-ik \cdot x} \diff x
\end{equation} 
denotes the Fourier transform of $f(x)$. Since we exclusively deal with $\RR^3$, we  often write $\Lp{p}$ and $\Sob{s}$ instead of $\Lp{p}(\RR^3)$ and $\Sob{s}(\RR^3)$ in what follows. A further abbreviation we use is given by
\begin{equation}
\int_{\RR^3} f \diff x := \int_{\RR^3} f(x) \diff x.
\end{equation}    
We equip $\Lp{2}(\RR^3)$ with a complex inner product, $\inner{\cdot}{\cdot}$, defined as 
\begin{equation} 
\inner{f}{g} := \int_{\RR^3} \bar{f} g \diff x.
\end{equation} 
Operator inequalities (in the sense of quadratic forms) are denoted by  $A \leq B$, which means that $\inner{\psi}{A \psi} \leq \inner{\psi}{B \psi}$ holds for all $\psi \in \set{D}(|A|^{1/2}) \subseteq \set{D}(|B|^{1/2})$, where $A$ and $B$ are self-adjoint operators on $\Lp{2}(\RR^3)$ with domains $\set{D}(A)$ and $\set{D}(B)$, respectively.

\section{Existence of Boosted Ground States}
\label{sec-existence}

We consider the following minimization problem
\begin{equation} \label{def-EvN}
E_v(N) := \inf \big \{ \Ev(\psi) : \psi \in \Hhalf(\RR^3), \; \Nfun(\psi) = N \big \},
\end{equation}
where $\Nfun(\psi)$ is defined in (\ref{def-Nfun}), and $N > 0$, $v \in \RR^3$, with $|v| < 1$, denote given parameters. Furthermore, we set
\begin{equation} \label{def-Ev}
\Ev(\psi) := \frac{1}{2} \innerb{\psi}{\Tm \psi} + \frac{i}{2} \innerb{\psi}{(\vel \cdot \nabla) \psi} - \frac{1}{4} \Vpot{\psi} .
\end{equation}
Any minimizer, $\sol_v \in \Hhalf(\RR^3)$, for (\ref{def-EvN}) has to satisfy the corresponding Euler-Lagrange equation given by
\begin{equation} \label{eq-boost1}
\Tm \sol_v + i (v \cdot \nabla) \sol_v - \big ( \frac{1}{|x|} \ast |\sol_v|^2 \big ) \sol_v = - \mu \sol_v,
\end{equation}
with some Lagrange multiplier, $-\mu \in \RR$, where this sign convention turns out to be convenient for our analysis. In what follows, we refer to such minimizers, $\sol_v$, for (\ref{def-EvN}) as {\em boosted ground states}, since they give rise to moving solitary waves
\begin{equation}
\psi(t,x) = e^{it\mu} \sol_v(x-vt),
\end{equation}
for (\ref{eq-hartree}) with travelling speed $v \in \RR^3$ with $|v| < 1$.

Concerning existence of boosted ground states, we have the following theorem which generalizes a result derived in \cite{Lieb+Yau1987} for minimizers of (\ref{def-EvN}) with $v=0$.

\begin{theorem} \label{th-existence}
Suppose that $m > 0$, $\vel \in \RR^3$, and $|\vel| < 1$. Then there exists a positive constant $\Nc(v)$ depending only on $v$ such that the following holds.
\begin{enumerate}
\item[i)] For $0 < N < \Nc(v)$, problem (\ref{def-EvN}) has a minimizer, $\sol_v \in \Hhalf(\RR^3)$, and it satisfies (\ref{eq-boost1}), for some $\mu \in \RR$. Moreover, every minimizing sequence, $(\psi_n)$, for (\ref{def-EvN}) with $0 < N < \Nc(v)$ is relatively compact in $\Hhalf(\RR^3)$ up to translations, \ie there exists a sequence, $(y_k)$, in $\RR^3$ and a subsequence, $(\psi_{n_k})$, such that $\psi_{n_k}( \cdot + y_k) \rightarrow \sol_v$ strongly in $\Hhalf(\RR^3)$ as $k \rightarrow \infty$, where $\sol_v$ is some minimizer for (\ref{def-EvN}).
\item[ii)] For $N \geq \Nc(v)$, no minimizer exists for problem (\ref{def-EvN}), even though $E_v(N)=-\frac{1}{2}mN$ is finite for $N = \Nc(v)$.
\end{enumerate}
\end{theorem}

\medskip
\noindent
{\bf Remarks.}  1) It has been proved in \cite{Lieb+Yau1987} that (\ref{def-EvN}) for $v=0$ has a spherically symmetric minimizer, which can be chosen to be real-valued and nonnegative. But the proof given in \cite{Lieb+Yau1987} crucially relies on symmetric rearrangement arguments that allow to restrict to radial functions in this special case. For $v \neq 0$, such methods cannot be used and a general discussion of (\ref{def-EvN}) needs a fundamental change of methods. Fortunately, it turns out that the concentration-compactness method introduced by P.-L.~Lions in \cite{Lions1984a} is tailor-made for studying (\ref{def-EvN}). To prove Theorem \ref{th-existence}, we shall therefore proceed along the lines of \cite{Lions1984a}. But --- due to the presence of the pseudo-differential operator $\sqrt{-\Delta + m^2}$ in (\ref{def-Ev}) --- some technical modifications have to be taken into account and they are worked out in detail in App.~A.

2) A corresponding existence result for boosted ground states can also be derived when $-1/|x|$ in (\ref{def-Ev}) is replaced by some other attractive two-body potential, \eg a Yukawa type potential $\Phi(x) = - e^{-\mu |x|}/|x|$ with $\mu > 0$. But then a minimal $L^2$-norm of minimizers has to be required, \ie the condition $N > N_*(v;\Phi)$ enters for some $N_*(v;\Phi) > 0$.

\subsection{Setting up the Variational Calculus}
\label{sec-setup-var-calc}

Before we turn to the proof of Theorem \ref{th-existence}, we collect and prove some preliminary results.

First one easily verifies that $\Ev(\psi)$ is real-valued (using, for instance, Plancherel's theorem for the first two terms in (\ref{def-Ev})). Moreover, the inequality 
\begin{equation} \label{ineq-Sopt}
\Vpot{\psi} \leq S_v \innerb{\psi}{(\T + i v \cdot \nabla) \psi} \innerb{\psi}{\psi},
\end{equation}
which is proven in App.~\ref{sec-app-best}, ensures that $\Ev(\psi)$ is well-defined on $\Hhalf(\RR^3)$. As stated in Lemma \ref{lem-constant}, inequality (\ref{ineq-Sopt}) has an optimizer, $Q_v \not \equiv 0$, for $|v| < 1$, which yields the best constant, $S_v$, in terms of
\begin{equation}
S_v = \frac{2}{ \inner{Q_v}{Q_v} } .
\end{equation}
Correspondingly, we introduce the constant, $\Nc(v)$, by 
\begin{equation} \label{def-Ncv}
\Nc(v) := \frac{2}{S_v} .
\end{equation} 
By Lemma \ref{lem-constant}, we also have the bounds
\begin{equation}
\Nc \geq \Nc(v) \geq (1-|v|) \Nc,
\end{equation}
where $\Nc(v=0) = \Nc > 4/\pi$ is, of course, the same constant that appeared in Sect.~1. 

We now state our first auxiliary result for (\ref{def-EvN}).

\begin{lemma} \label{lem-Ev-lower}
Suppose that $m \geq 0$, $v \in \RR^3$, and $|v| < 1$. Then the following inequality holds
\begin{equation} \label{ineq-Ev-apriori}
2 \Ev(\psi) \geq  \big ( 1 - \frac{N}{\Nc(v)} \big ) \innerb{\psi}{\big ( \T +i v \cdot \nabla \big ) \psi} - mN 
\end{equation}
for all $\psi \in \Hhalf(\RR^3)$ with $\Nfun(\psi) = N$. Here $\Nc(v)$ is the constant introduced in (\ref{def-Ncv}) above.

Moreover, we have that $E_v(N) \geq -\frac{1}{2} mN$ for $0 < N \leq \Nc(v)$ and $E_v(N) = -\infty$ for $N > \Nc(v)$. Finally, any minimizing sequence for problem (\ref{def-EvN}) is bounded in $\Hhalf(\RR^3)$ whenever $0 < N < \Nc(v)$. 
\end{lemma}

\begin{proof}[Proof of Lemma \ref{lem-Ev-lower}]
Let the assumption on $m$ and $v$ stated above be satisfied. Estimate (\ref{ineq-Ev-apriori}) is derived by noting that  $\sqrt{-\Delta + m^2} \geq \T$ and using inequality (\ref{ineq-Sopt}) together with the definition of $\Nc(v)$ in (\ref{def-Ncv}). Furthermore, that $E_v(N) \geq - \frac{1}{2} mN$ for $N \leq \Nc(v)$ is a consequence of (\ref{ineq-Ev-apriori}) itself. To see that $E_v(N) = -\infty$ when $N > \Nc(v)$, we recall from Lemma \ref{lem-constant} that there exists an optimizer, $Q_v \in \Hhalf(\RR^3)$, with $\Nfun(Q_v) = \Nc(v)$, for inequality (\ref{ineq-Sopt}). Using that $Q_v$ turns (\ref{ineq-Sopt}) into an equality and noticing that $\sqrt{-\Delta +m^2} - m \leq \T$ holds, a short calculation yields
\begin{equation}
E_v(N) \leq \Ev(\lambda Q_v) \Big |_{m=0} = -\frac{ \lambda^2(\lambda^2-1)}{4} \Vpot{Q_v} . 
\end{equation}  
For $N > \Nc(v)$, we can choose $\lambda > 1$ which implies that the right-hand side is strictly negative and, in addition, by $L^2$-norm preserving rescalings, $Q_v(x) \mapsto a^{3/2}Q_v(ax)$ with $a > 0$, we find that
\begin{equation}
E_v(N) \leq \Ev(\lambda a^{3/2}Q_v(a \cdot) ) \Big |_{m=0} = a \Ev(\lambda Q_v) \Big |_{m=0} \rightarrow -\infty, \quad \mbox{with $\lambda > 1$ as $a \rightarrow \infty$}.
\end{equation}
Thus, we deduce that $E_v(N) = -\infty$ holds whenever $N > \Nc(v)$.

To see the $\Hhalf(\RR^3)$-boundedness of any minimizing sequence, $(\psi_n)$, with $0 < N < \Nc(v)$, we note that $\T + i v\cdot \nabla \geq (1-|v|) \T$ holds. Hence we see that $\sup_{n} \inner{\psi_n}{\T \psi_n} \leq C < \infty$, thanks to (\ref{ineq-Ev-apriori}). This completes the proof of Lemma \ref{lem-Ev-lower}.  \end{proof}

As a next step, we derive an upper bound for $E_v(N)$, which is given by the nonrelativistic ground state energy, $E^{\set{nr}}_v(N)$, defined below. Here the positivity of the mass parameter, $m >0$, is essential for deriving the following estimate.

\begin{lemma} \label{lem-upper}
Suppose that $m > 0$, $v \in \RR^3$, and $|v| < 1$. Then we have that
\begin{equation} \label{ineq-Enr}
E_v(N) \leq -\frac{1}{2} \big ( 1 - \sqrt{1-v^2} \big ) mN + E^{\set{nr}}_v(N),
\end{equation}
where $E^{\set{nr}}_v(N)$ is given by 
\begin{equation}
E^{\set{nr}}_v(N) := \inf \big \{ \mathcal{E}^{\set{nr}}_v(\psi)   : \psi \in \Hone(\RR^3), \; \mathcal{N}(\psi) = N \big \}  ,
\end{equation}
\begin{equation}
\mathcal{E}^{\set{nr}}_v(\psi) := \frac{\sqrt{1-v^2}}{4 m} \int_{\RR^3} | \nabla \psi |^2 \diff x - \frac{1}{4} \Vpot{\psi} .
\end{equation} 
\end{lemma}

\begin{proof}[Proof of Lemma \ref{lem-upper}]
To prove (\ref{ineq-Enr}), we pick a spherically symmetric function, $\phi \in \Hone(\RR^3)$ with $\Nfun(\phi) = N$, and we introduce the one-parameter family
\begin{equation}
\phi_\lambda(x) := e^{i \lambda v \cdot x} \phi(x) = e^{i  \lambda |v|z} \phi(x), \quad \mbox{with $\lambda > 0$.}
\end{equation} 
Here and in what follows, we assume (without loss of generality) that $v$ is parallel to the $z$-axis, \ie $v = |v| e_z$. One checks that
\begin{equation}
\frac{i}{2} \innerb{ \phi_\lambda}{(v \cdot \nabla) \phi_\lambda} = -\frac{\lambda v^2}{2} N,
\end{equation}
using the fact that $\inner{\phi}{\nabla \phi} = 0$ holds, by spherically symmetry of $\phi(x)$. Hence, we find that
\begin{align}
\Ev(\phi_\lambda) & = \frac{1}{2} \innerb{\phi_\lambda}{\Tm \phi_\lambda} + \frac{i |v|}{2} \innerb{\phi_\lambda}{ \partial_z \phi_\lambda} - \frac{1}{4} \Vpot{\phi_\lambda} \nonumber \\
& = \frac{1}{2} \left ( \innerb{\phi_\lambda}{\Tm \phi_\lambda} - v^2 \lambda N \right ) - \frac{1}{4} \Vpot{\phi} \nonumber  \\
& =: A + B . \label{ineq-Ev-boost}
\end{align} 
To estimate $A$ in (\ref{ineq-Ev-boost}), we recall the operator inequality
\begin{equation}
\sqrt{-\Delta+m^2} \leq \frac{1}{2 \lambda}  ( -\Delta + m^2 + \lambda^2  ),
\end{equation}
which follows from the elementary inequality $2 |a| |b| \leq a^2 + b^2$. Thus, we are led to
\begin{align}
A & \leq \frac{1}{4 \lambda} \innerb{\phi_\lambda}{ ( -\Delta + m^2 + \lambda^2 ) \phi_\lambda} - \frac{1}{2} mN - \frac{1}{2} v^2 \lambda N  \nonumber \\
& = \frac{1}{4 \lambda} \big ( \lambda^2 v^2 N + \inner{\phi}{-\Delta \phi} + (m^2 + \lambda^2) N \big )  - \frac{1}{2}mN - \frac{1}{2} v^2 \lambda N . \label{ineq-upper}
\end{align}
By minimizing the upper bound (\ref{ineq-upper}) with respect to $\lambda > 0$, which is a matter of elementary calculations, we obtain with $\lambda_* = m/\sqrt{1-v^2}$ the estimate
\begin{align}
\Ev(\phi_{\lambda_*}) & \leq -\frac{1}{2} \big ( 1 - \sqrt{1-v^2} \big ) mN + \frac{\sqrt{1-v^2}}{4m} \inner{\phi}{-\Delta \phi} - \frac{1}{4} \Vpot{\phi}  \nonumber \\
& = -\frac{1}{2} \big ( 1 - \sqrt{1-v^2} \big ) mN + \mathcal{E}^{\set{nr}}_v(\phi ) . 
\end{align}
Next, we remark that $\mathcal{E}^{\mathrm{nr}}_v(\psi)$ is the energy functional for the non-relativistic boson star problem with mass parameter $m_v= m/\sqrt{1-v^2}$. Indeed, it is known from \cite{Lieb1977} that $\mathcal{E}^{\mathrm{nr}}_v(\psi)$ subject to $\Nfun(\psi) = N$ has a spherically symmetric minimizer, $\phi_* \in \Hone(\RR^3)$, with 
\begin{equation} \label{eq-Enr}
E^{\set{nr}}_v(N) = \mathcal{E}^{\mathrm{nr}}_v(\phi_*) < 0,
\end{equation}
which completes the proof of Lemma \ref{lem-upper}. \end{proof}

By making use of Lemma \ref{lem-upper}, we show that the function $E_v(N)$ satisfies a strict subadditivity condition. This is essential to the discussion of (\ref{def-EvN}) when using concentration-compactness-type methods. 

\begin{lemma} \label{lem-sab}
Suppose that $m > 0$, $\vel \in \RR^3$, and $|\vel| < 1$. Then $E_v(N)$ satisfies the strict subadditivity condition
\begin{equation} \label{ineq-sab}
E_v(N) < E_v(\alpha) + E_v(N- \alpha)
\end{equation}
whenever $0 < N < \Nc(v)$ and $0 < \alpha < N$. Here $\Nc(v)$ is the constant of Lemma \ref{lem-Ev-lower}.

Moreover, the function $E_v(N)$ is strictly decreasing and strictly concave in $N$, where $0 < N < \Nc(v)$.
\end{lemma}

\medskip
\noindent
{\bf Remarks.} 1) Condition $m > 0$ is necessary for (\ref{ineq-sab}) to hold. To see this, note that if $m=0$ then $\Ev(\psi_\lambda) = \lambda \Ev(\psi)$ holds, where $\psi_\lambda = \lambda^{3/2} \psi(\lambda x)$ and $\lambda > 0$. This leads to the conclusion that $E_v(N)$ is either $0$ or $-\infty$ when $m=0$.

2) The fact that $E_v(N)$ is strictly concave will be needed in our companion paper \cite{FJL2005II} when making use of the symplectic structure associated with the Hamiltonian PDE (\ref{eq-hartree}). More precisely, the strict concavity of $E_v(N)$ will enable us to prove the nondegeneracy of the symplectic form restricted to the manifold of solitary waves.  

\begin{proof}[Proof of Lemma \ref{lem-sab}]  
By Lemma \ref{lem-upper} and the fact that $E^{\set{nr}}_v(N) \leq E^{\set{nr}}_{v=0}(N) < 0$ holds, by (\ref{eq-Enr}), we deduce that 
\begin{equation} \label{ineq-EvN2}
E_v(N) < -\frac{1}{2} \big ( 1 - \sqrt{1-v^2} \big ) m N ,
\end{equation}
Next, we notice the following scaling behavior
\begin{equation} \label{eq-evn}
E_v(N) =  N e_v(N),
\end{equation}
where 
\begin{align}
e_v(N)  := \inf_{\psi \in \Hhalf, \| \psi \|_2^2 = 1} & \Big \{ \frac{1}{2} \innerb{\psi}{\Tm \psi} + \frac{i}{2} \innerb{\psi}{(v \cdot \nabla) \psi} \nonumber \\
& \quad - \frac{N}{4} \Vpot{\psi} \Big \} .
\end{align}
This shows that $e_v(N)$ is strictly decreasing, provided that we know that we may restrict the infimum to elements such that
\begin{equation} \label{ineq-Vlo}
\Vpot{\psi} \geq c > 0
\end{equation}
holds for some $c$. Suppose now that (\ref{ineq-Vlo}) were not true. Then there exists a minimizing sequence, $(\psi_n)$, such that
\begin{equation}
\Vpot{\psi_n} \rightarrow 0, \quad \mbox{as $n \rightarrow \infty$}.
\end{equation}
But on account of the fact that (cf.~App.~C)
\begin{equation}
\innerb{\psi}{\Tm \psi} + i  \innerb{\psi}{(v \cdot \nabla) \psi} \geq - \big ( 1-\sqrt{1-v^2} \big ) m N,
\end{equation}
we conclude that 
\begin{equation} \label{ineq-false}
E_v(N) = N e_v(N) \geq -\frac{1}{2} \big ( 1 - \sqrt{1-v^2} \big ) mN ,
\end{equation}
which contradicts (\ref{ineq-EvN2}). Thus $e_v(N)$ is strictly decreasing. Returning to (\ref{eq-evn}) and noting that $e_v(N) < 0$ holds, by (\ref{ineq-EvN2}), we deduce that
\begin{equation}
E_v(\vartheta N) < \vartheta E_v(N), \quad \mbox{for $\vartheta > 1$ and $0 < N < \Nc(v)$.}
\end{equation}
By an argument presented in \cite{Lions1984a}, this inequality leads to the strict subadditivity condition (\ref{ineq-sab}). 

Finally, we show that $E_v(N)$ is strictly decreasing and strictly concave on the interval $(0,\Nc(v))$. To see that $E_v(N)  = N e_v(N)$ is strictly decreasing, we notice that $e_v(N)$ is strictly decreasing and negative. Furthermore, we remark that $e_v(N) = \inf \{ \mbox{linear functions in $N$} \}$ has to be a concave function. Therefore it follows that $E_v(N) = N e_v(N)$  is a strictly concave, since the left- and right-derivatives, $D^\pm E_v(N)$, exist and are found to be strictly decreasing, by using that $e_v(N)$ is concave and strictly decreasing. \end{proof}

\subsection{Proof of Theorem \ref{th-existence}}
\label{sec-proof-existence}

We now come to the proof of Theorem \ref{th-existence} and we suppose that $m > 0$, $v \in \RR^3$, and $|v| < 1$ holds.

\subsubsection*{Proof of Part i)}

Let us assume that
\begin{equation} 
0 < N < \Nc(v), 
\end{equation}
where $\Nc(v)$ is the constant defined in (\ref{def-Ncv}). Furthermore, let $(\psi_n)$ be a minimizing sequence for (\ref{def-EvN}), \ie
\begin{equation}
\lim_{n \rightarrow \infty} \Ev(\psi_n) = E_v(N), \quad \mbox{with $\psi_n \in \Hhalf(\RR^3)$ and $\Nfun(\psi_n) = N$ for all $n \geq 0$}.
\end{equation} 
By Lemma \ref{lem-Ev-lower}, we have that $E_v(N) > -\infty$ and that $(\psi_n)$ is a bounded sequence in $\Hhalf(\RR^3)$. We now apply the the following concentration-compactness lemma.

\begin{lemma} \label{lem-con-com}
Let $( \psi_n )$ be a bounded sequence in $\Hhalf(\RR^3)$ with $\Nfun(\psi_n) = \int_{\RR^3} |\psi_n|^2 \diff x = N$ for all $n \geq 0$. Then there exists a subsequence, $( \psi_{n_k} )$, satisfying one of the three following properties.
\begin{enumerate}
\item[i)] \underline{Compactness:} There exists a sequence, $(y_k)$, in $\RR^3$ such that, for every $\epsilon >0$, there exists $0 < R < \infty$ with
\begin{equation}
 \int_{|x-y_k|<R} | \psi_{n_k} |^2 \diff x \geq N - \epsilon . 
\end{equation}
\item[ii)] \underline{Vanishing:} 
\[  \lim_{k \rightarrow \infty} \sup_{y \in \RR^3} \int_{|x-y|<R} |\psi_{n_k} |^2 \diff x = 0, \quad \mbox{for all $R > 0$.} \]
\item[iii)] \underline{Dichotomy:} There exists $\alpha \in (0,N)$ such that, for every $\epsilon > 0$, there exist two bounded sequences, $(\psi^1_k)$ and $(\psi^2_k)$, in $\Hhalf(\RR^3)$ and $k_0 \geq 0$ such that, for all $k \geq k_0$, the following properties hold 
\begin{equation} \label{ineq-con-com-Lp}
\big \| \psi_{n_k} - ( \psi^1_k + \psi^2_k) \big \|_p \leq \delta_p(\epsilon), \quad \mbox{for $2 \leq p < 3$},
\end{equation}
with $\delta_p(\epsilon) \rightarrow 0$ as $\epsilon \rightarrow 0$, and
\begin{equation} \label{ineq-con-com-mass}
\Big | \int_{\RR^3} | \psi^1_k |^2 \diff x - \alpha \Big | \leq \epsilon \quad \mbox{and} \quad \Big | \int_{\RR^3} | \psi^2_k |^2 \diff x - (N-\alpha) \Big | \leq \epsilon,
\end{equation}
\begin{equation} \label{ineq-con-com-supp}
\mathrm{dist} \, ( \mathrm{supp} \, \psi^1_k, \, \mathrm{supp} \, \psi^2_k ) \rightarrow \infty, \quad \mbox{as $k \rightarrow \infty$}.
\end{equation}
Moreover, we have that
\begin{equation} \label{ineq-con-com-liminf-T}  
\liminf_{k \rightarrow \infty} \Big ( \inner{\psi_{n_k}}{T \psi_{n_k}} - \innerb{\psi^1_k}{T \psi^1_k} - \innerb{\psi^2_k}{T \psi^2_k} \Big ) \geq -C(\epsilon), 
\end{equation}
where $C(\epsilon) \rightarrow 0$ as $\epsilon \rightarrow 0$ and $T := \Tm + i (v \cdot \nabla)$ with $m \geq 0$ and $v \in \RR^3$.
\end{enumerate}
\end{lemma}

\medskip
\noindent
{\bf Remark.} We refer to App.~A for the proof of Lemma \ref{lem-con-com}. Part i) and ii) are standard, but part iii) requires some technical arguments, due to the presence of the pseudo-differential operator $T$.

\medskip
\noindent
Invoking Lemma \ref{lem-con-com}, we conclude that a suitable subsequence, $(\psi_{n_k})$, satisfies either i), ii), or iii). We rule out ii) and iii) as follows. 

Suppose that $(\psi_{n_k})$ exhibits property ii). Then we conclude that
\begin{equation}
\lim_{k \rightarrow \infty} \Vpot{\psi_{n_k}} = 0,
\end{equation}
by Lemma \ref{lem-vanish}. But as shown in the proof of Lemma \ref{lem-sab}, this implies
\begin{equation}
E_v(N) \geq - \frac{1}{2} \big ( 1- \sqrt{1-v^2} \big ) mN,
\end{equation}
which contradicts (\ref{ineq-EvN2}). Hence ii) cannot occur.

Let us suppose that iii) is true for $(\psi_{n_k})$. Then there exists $\alpha \in (0,N)$ such that, for every $\epsilon > 0$, there are two bounded sequences, $(\psi^1_k)$ and $(\psi^2_k)$, with
\begin{equation} \label{ineq-psi12}
 \alpha - \epsilon \leq \Nfun(\psi^1_k) \leq \alpha + \epsilon, \quad (N-\alpha) - \epsilon \leq \Nfun(\psi^2_k) \leq (N-\alpha) + \epsilon,
\end{equation}
for $k$ sufficiently large. Moreover, inequality (\ref{ineq-con-com-liminf-T}) and Lemma \ref{lem-dicho} allow us to deduce that
\begin{equation}
E_v(N) = \lim_{k \rightarrow \infty} \Ev(\psi_{n_k}) \geq \liminf_{k \rightarrow \infty} \Ev(\psi^1_k) + \liminf_{k \rightarrow \infty} \Ev(\psi^2_k) - r(\epsilon),
\end{equation}
where $r(\epsilon) \rightarrow 0$ as $\epsilon \rightarrow 0$. Since $(\psi^1_k)$ and $(\psi^2_k)$ satisfy (\ref{ineq-psi12}), we infer 
\begin{equation}
E_v(N) \geq E_v(\alpha + \epsilon) + E_v(N-\alpha + \epsilon) - r(\epsilon), 
\end{equation}
using that $E_v(N)$ is decreasing in $N$. Passing to the limit $\epsilon \rightarrow 0$ and by continuity of $E_v(N)$ in $N$ (recall that $E_v(N)$ is concave function on an open set), we deduce that
\begin{equation}
E_v(N) \geq E_v( \alpha) + E_v(N-\alpha)
\end{equation}
holds for some $0 < \alpha < N$. This contradicts the strict subadditivity condition (\ref{ineq-sab}) stated in Lemma \ref{lem-sab}. Therefore iii) is ruled out.

By the discussion so far, we conclude that there exists a subsequence, $(\psi_{n_k})$, such that i) of Lemma \ref{lem-con-com} is true for some sequence $(y_k)$ in $\RR^3$. Let us now define the sequence
\begin{equation}
\widetilde{\psi}_k := \psi_{n_k}(\cdot + y_k) .
\end{equation}
Since $(\widetilde{\psi}_k)$ is a bounded sequence in $\Hhalf(\RR^3)$, we can pass to a subsequence, still denoted by $(\widetilde{\psi}_k)$, such that $(\widetilde{\psi}_k)$ converges weakly in $\Hhalf(\RR^3)$ to some $\sol_v \in \Hhalf(\RR^3)$ as $k \rightarrow \infty$. Moreover, we have that $\widetilde{\psi}_k \rightarrow \sol_v$ strongly in $\Lploc{p}(\RR^3)$ as $k \rightarrow \infty$, for $2 \leq p < 3$, thanks to a Rellich-type theorem for $\Hhalf(\RR^3)$ (see, \eg \cite[Theorem 8.6]{Lieb+LossII} for this). But on account of the fact
\begin{equation}
\int_{|x| < R} | \widetilde{\psi}_k |^2 \, \dd x \geq N - \epsilon,
\end{equation} 
for every $\epsilon > 0$ and suitable $R = R(\epsilon) < \infty$, we conclude that 
\begin{equation} \label{eq-strconv}
\mbox{$\widetilde{\psi}_k \rightarrow \sol_v$ strongly in $\Lp{p}(\RR^3)$ as $k \rightarrow \infty$, for $2 \leq p < 3$. }
\end{equation}
Next, by the Hardy-Littlewood-Sobolev inequality, we deduce that
\begin{align*}
\Big | \Vpot{\widetilde{\psi}_k} - \Vpot{\sol_v} \Big | & \leq C \big \| | \widetilde{\psi}_k|^2 + | \sol |^2 \big \|_{6/5} \big \| | \widetilde{\psi}_k|^2 - | \sol_v |^2 \big \|_{6/5} \nonumber \\
& \leq C \big ( \| \widetilde{\psi}_k \|^3_{12/5} + \| \sol_v \|_{12/5}^3 \big ) \|  \widetilde{\psi}_k - \sol_v \|_{12/5}.
\end{align*}
From (\ref{eq-strconv}), we have that $\widetilde{\psi}_k$ converges strongly to $\sol_v$ in $\Lp{12/5}(\RR^3)$, as $k \rightarrow \infty$, and therefore 
\begin{equation}
\lim_{k \rightarrow \infty} \Vpot{\widetilde{\psi}_k} = \Vpot{\sol_v} .
\end{equation}
Moreover, we have that
\begin{equation} \label{ineq-wlsc-min}
E_v(N) = \lim_{k \rightarrow \infty} \Ev(\widetilde{\psi}_k) \geq \Ev(\sol_v) \geq E_v(N),
\end{equation}
since the functional 
\begin{equation} 
 \mathcal{T}(\psi) := \innerb{\psi}{ \Tm \psi } + i \innerb{ \psi}{ (v \cdot \nabla) \psi}, 
\end{equation}
is weakly lower semicontinuous on $\Hhalf(\RR^3)$, see Lemma \ref{lem-wlsc} in App.~A. Thus, we have proved that $\sol_v \in \Hhalf(\RR^3)$ is a minimizer for (\ref{def-EvN}), \ie we have $E_v(N) = \Ev(\sol_v)$ and $\Nfun(\sol_v)  = N$.

Next, we address the relative compactness of minimizing sequences in $\Hhalf(\RR^3)$ (up to translations). To do so, we notice that there has to be equality in (\ref{ineq-wlsc-min}), which leads to $ \lim_{k \rightarrow \infty} \mathcal{T}(\widetilde{\psi}_k) = \mathcal{T}(\sol_v)$. By Lemma \ref{lem-wlsc}, this fact implies a posteriori that 
\begin{equation}
\mbox{$\widetilde{\psi}_{k} \rightarrow \sol_v$ strongly in $\Hhalf(\RR^3)$ as $k \rightarrow \infty$},
\end{equation}
which completes the proof of part i) of Theorem \ref{th-existence}.

\subsubsection*{Proof of Part ii)}

To complete the proof of Theorem \ref{th-existence}, we address its part ii). Clearly, no minimizer exists if $N > \Nc(v)$, since in this case we have that $E_v(N) = -\infty$, by Lemma \ref{lem-Ev-lower}. 

Next, we show that $E_v(N) = -\frac{1}{2}mN$ holds if $N= \Nc(v)$, which can be seen as follows. We take an optimizer, $Q_v \in \Hhalf(\RR^3)$, for inequality (\ref{ineq-Sopt}); see Lemma \ref{lem-constant} and recall that $\Nfun(Q_v) = \Nc(v)$. Then 
\begin{equation} \label{ineq-EvN-above}
E_v(N) \leq \Ev(Q^{(\lambda)}_v) = \frac{1}{2} \innerb{Q^{(\lambda)}_v}{\big ( \sqrt{-\Delta + m^2} - \sqrt{-\Delta} \big ) Q^{(\lambda)}_v} - \frac{1}{2} m N \quad \mbox{for $N = \Nc(v)$},
\end{equation} 
where $Q^{(\lambda)}_v(x) := \lambda^{3/2} Q_v(\lambda x)$ with $\lambda > 0$, so that $\Nfun(Q^{(\lambda)}_v) = \Nfun(Q_v) = \Nc(v)$. Using Plancherel's theorem and by dominated convergence, we deduce that
\begin{align}
\innerb{Q^{(\lambda)}_v}{\big ( \sqrt{-\Delta + m^2} - \sqrt{-\Delta} \big ) Q^{(\lambda)}_v} & = \int_{\RR^3} |\widehat{Q}_v(k)|^2 \big ( \sqrt{ \lambda^2 k^2 + m^2} - \lambda |k| \big ) \diff k \nonumber  \\
 & \rightarrow 0 \quad \mbox{as $\lambda \rightarrow \infty$}.
\end{align}
Thus, we conclude that $E_v(N) \leq -\frac{1}{2} mN$ for $N = \Nc(v)$. In combination with the estimate $E_v(N) \geq -\frac{1}{2} mN$ for $N \leq \Nc(v)$ taken from Lemma \ref{lem-Ev-lower}, this shows that 
\begin{equation}
E_v(N) = - \frac{1}{2} mN \quad \mbox{for $N = \Nc(v)$.}
\end{equation} 

Finally, we prove that there does not exist a minimizer for (\ref{def-EvN}) with $N = \Nc(v)$. We argue by contradiction as follows. Suppose that $\sol_v \in \Hhalf(\RR^3)$ is a minimizer for (\ref{def-EvN}) with $N = \Nc(v)$. Using the strict inequality $\inner{\sol_v}{\sqrt{-\Delta + m^2} \sol_v} > \inner{\sol_v}{\T \, \sol_v}$, for $m>0$ and $\sol_v \in \Hhalf(\RR^3)$, $\sol_v \not \equiv 0$, we obtain  
\begin{equation}
-\frac{1}{2} mN = \Ev(\sol_v) \big |_{m > 0} > \Ev(\sol_v) \big |_{m=0} - \frac{1}{2} mN \geq -\frac{1}{2} mN,
\end{equation} 
which is a contradiction. Here we use Lemma \ref{lem-Ev-lower} in order to estimate $\Ev(\sol_v) |_{m=0} \geq 0$ for $\Nfun(\sol_v)  = \Nc(v)$. Hence no minimizer exists for (\ref{def-EvN}) if $N \geq \Nc(v)$. This completes the proof of Theorem \ref{th-existence}. \hfill $\qed$

\section{Orbital Stability}
\label{sec-stability}

The purpose of this section is to address {\em ``orbital stability''} of travelling solitary waves
\begin{equation} \label{eq-boost}
\psi(t,x) = e^{i t \mu} \sol_v(x-vt) ,
\end{equation}
where $\sol_v \in \Hhalf(\RR^3)$ is a boosted ground state. By the relative compactness of minimizing sequences (see Theorem \ref{th-existence}) and by using a general idea presented in \cite{Cazenave+Lions1982}, we are able to prove the following abstract stability result.
\begin{theorem} \label{th-stability}
Suppose that $m > 0$, $\vel \in \RR^3$, $|v| < 1$, and $0 < N < \Nc(v)$. Let $\set{S}_{v,N}$ denote the corresponding set of boosted ground states, \ie
\[
\set{S}_{v,N} := \big \{ \sol_v \in \Hhalf(\RR^3): \Ev(\sol_v) = E_v(N), \; \Nfun(\sol_v) = N \big \} ,
\]
which is non-empty by Theorem \ref{th-existence}.

Then the solitary waves given in (\ref{eq-boost}), with $\sol_v \in \set{S}_{v,N}$, are stable in the following sense. For every $\epsilon > 0$, there exists $\delta  > 0$ such that 
\[ 
\inf_{\sol_v \in \set{S}_{v,N}} \nrmHhalf{ \psi_0 - \sol_v } \leq \delta \quad \mbox{implies that} \quad  \sup_{t \geq 0} \inf_{\sol_v \in \set{S}_{v,N}} \nrmHhalf{ \psi(t) - \sol_v } \leq \epsilon.
\]
Here $\psi(t)$ denotes the solution of (\ref{eq-hartree}) with initial condition $\psi_0 \in \Hhalf(\RR^3)$.
\end{theorem}

\begin{proof}[Proof of Theorem \ref{th-stability}]
Let $m$ and $\vel$ satisfy the given assumptions. Since we have $N < \Nc(v) \leq \Nc$, we can choose $\delta > 0$ sufficiently small such that $\inf_{\phi \in \set{S}_{v,N}} \nrmHhalf{\psi_0 - \phi} \leq \delta$ guarantees that $\Nfun(\psi_0) < \Nc$. By the global well-posedness result for (\ref{eq-hartree}) derived in \cite{Lenzmann2005LWP}, we have that the corresponding solution, $\psi(t)$, exists for all times $t \geq 0$. Thus, taking $\sup_{t \geq 0}$ is well-defined.

Let us now assume that orbital stability (in the sense defined above) does not hold. Then there exists a sequence on initial data, $(\psi_n(0))$, in $\Hhalf(\RR^3)$ with
\begin{equation} \label{eq-instab1}
\inf_{\sol \in \set{S}_{v,N}} \nrmHhalf{\psi_n(0) - \sol} \rightarrow 0, \quad \mbox{as $n \rightarrow \infty$},
\end{equation} 
and some $\epsilon > 0$ such that
\begin{equation} \label{eq-instab2}
\inf_{\sol \in \set{S}_{v,N}} \nrmHhalf{\psi_n(t_n) - \sol} > \epsilon,  \quad \mbox{for all $n \geq 0$},
\end{equation}
for a suitable sequence of times $(t_n)$. Note that (\ref{eq-instab1}) implies that $\Nfun(\psi_n(0)) \rightarrow N$ as $n \rightarrow \infty$. Since $N < \Nc$ by assumption, we can assume --- without loss of generality --- that $\Nfun( \psi_n(0)) < \Nc$ holds for all $n \geq 0$, which guarantees (see above) that the corresponding solution, $\psi_n(t)$, exists globally in time.

Next, we consider the sequence, $(\beta_n)$, in $\Hhalf(\RR^3)$ that is given by 
\begin{equation}
\beta_n := \psi_n(t_n) .
\end{equation}
By conservation of $\Nfun(\psi(t))$ and of $\Ev(\psi(t))$, whose proof can be done along the lines of \cite{Lenzmann2005LWP} for the conservation of $\mathcal{E}(\psi(t))$, we have that $\Nfun(\beta_n) = \Nfun(\psi_n(0))$ and $\Ev(\beta_n) = \Ev(\psi_n(0))$, which, by (\ref{eq-instab1}), implies 
\begin{equation}
\lim_{n \rightarrow \infty} \Ev(\beta_n) = E_v(N) \quad \mbox{and} \quad \lim_{n \rightarrow \infty} \Nfun(\beta_n) = N.
\end{equation}
Defining the rescaled sequence  
\begin{equation}
\widetilde{\beta}_n := a_n \beta_n, \quad \mbox{where $a_n := \sqrt{N/\Nfun(\beta_n)}$},
\end{equation}
and using the fact $(\beta_n)$ has to be bounded in $\Hhalf(\RR^3)$, by virtue of Lemma \ref{lem-Ev-lower}, we infer that 
\begin{equation} \label{ineq-stab1}
\nrmHhalf{ \beta_n - \widetilde{\beta}_n } \leq C |1-a_n| \rightarrow 0, \quad \mbox{as $n \rightarrow \infty$}.
\end{equation}
By continuity of $\Ev : \Hhalf(\RR^3) \rightarrow \RR$, we deduce that
\begin{equation}
\lim_{n \rightarrow \infty} \Ev(\widetilde{\beta}_n) = E_v(N) \quad \mbox{and} \quad  \Nfun(\widetilde{\beta}_n) = N, \quad \mbox{for all $n \geq 0$.}
\end{equation}
Therefore $(\widetilde{\beta}_n)$ is a minimizing sequence for (\ref{def-EvN}) which, by Theorem \ref{th-existence} part i), has to contain a subsequence, $(\widetilde{\beta}_{n_k})$, that strongly converges in $\Hhalf(\RR^3)$ (up to translations) to some minimizer $\sol \in \set{S}_{v,N}$. In particular, inequality (\ref{eq-instab2}) cannot hold when $\beta_n=\psi_n(t_n)$ is replaced by $\widetilde{\beta}_n$. But in view of (\ref{ineq-stab1}), this conclusion is easily extended to the sequence $(\beta_n)$ itself. Thus, we are led to a contradiction and the proof of Theorem \ref{th-stability} is complete.\end{proof}

\section{Properties of Boosted Ground States}
\label{sec-properties}

Concerning fundamental properties of boosted ground states given by Theorem \ref{th-existence}, we have the following result. 

\begin{theorem} \label{th-reg-decay}
Let $m > 0$, $v \in \RR^3$, $|v| < 1$, and $0 < N < \Nc(v)$. Then every boosted ground state, $\sol_v \in \Hhalf(\RR^3)$, of problem (\ref{def-EvN}) satisfies the following properties. 
\begin{enumerate}
\item[i)] $\sol_v \in \Sob{s}(\RR^3)$ for all $s \geq 1/2$.
\item[ii)] The corresponding Lagrange multiplier satisfies $\mu > (1 - \sqrt{1-v^2}) m$. Moreover, we have pointwise exponential decay, \ie
\begin{equation}
|\sol_v(x)| \leq C e^{-\delta |x|}
\end{equation} 
holds for all $x \in \RR^3$, where $\delta > 0$ and $C > 0$ are suitable constants.
\item[iii)] For $v = 0$, the function $\sol_v(x)$ can be chosen to be radial, real-valued, and strictly positive.
\end{enumerate}
\end{theorem}

\medskip
\noindent
{\bf Remarks.} 1) By part i) and Sobolev embeddings, any boosted ground state is smooth: $\sol_v \in \C{\infty}(\RR^3)$. Moreover, we have that $\sol_v \in \Lp{1} \cap \Lp{\infty}$, due to part ii).   

2) Part iii) follows from the discussion presented in \cite{Lieb+Yau1987}, except for the strict positivity which we will show below.

3) For a more precise exponential decay estimate for $\sol_v(x)$, see Lemma \ref{lem-exp-decay} in App.~C.

\begin{proof}[Proof of Theorem \ref{th-reg-decay}]
Part i): We rewrite the Euler-Lagrange equation (\ref{eq-boost}) for $\sol_v$ as 
\begin{equation} \label{eq-reg1}
(H_0 + \lambda ) \sol_v = F(\sol_v) + (\lambda - \mu) \sol_v,
\end{equation}
for any $\lambda \in \RR$, where
\begin{equation}
H_0 := \Tm + i (v \cdot \nabla) , \quad F(\sol_v) := \big ( \frac{1}{|x|} \ast |\sol_v|^2 \big ) \sol_v .
\end{equation}
By \cite[Lemma 3]{Lenzmann2005LWP}, we have that $F : \Sob{s}(\RR^3) \rightarrow \Sob{s}(\RR^3)$ for all $s \geq 1/2$ ($F$ is indeed locally Lipschitz). Thus, the right-hand side in (\ref{eq-reg1}) belongs to $\Hhalf(\RR^3)$. Since $H_0$ is bounded from below, we can choose $\lambda > 0$ sufficiently large such that $(H_0 + \lambda)^{-1}$ exists. This leads to
\begin{equation} \label{eq-reg2}
\sol_v = (H_0 + \lambda)^{-1} \big [ F(\sol_v) + (\lambda - \mu) \sol_v \big ].
\end{equation}
Noting that $(H_0 + \lambda)^{-1} : \Sob{s}(\RR^3) \rightarrow \Sob{s+1}(\RR^3)$, we see that $\sol_v \in \Sob{3/2}(\RR^3)$. By repeating the argument, we conclude that $\sol_v \in  \Sob{s}(\RR^3)$ for all $s \geq 1/2$. This proves part i).

Part ii): The exponential decay follows from Lemma \ref{lem-exp-decay}, provided that the Lagrange multiplier, $-\mu$, satisfies
\begin{equation} \label{ineq-reg2}
- \mu < - \big (1 - \sqrt{1-v^2}  \big )  m ,
\end{equation}
which means that $-\mu$ lies strictly below the essential spectrum of $H_0$; see App.~C. To prove (\ref{ineq-reg2}), we multiply the Euler-Lagrange equation by $\overline{\sol}_v$ and integrate to obtain
\begin{equation}
2 E_v(N) - \frac{1}{2} \Vpot{\sol_v} = - \mu N .
\end{equation}
Using the upper bound (\ref{ineq-EvN2}) for $E_v(N)$, we conclude that
\begin{equation}
- \mu N < -\big ( 1 - \sqrt{1-v^2} \big ) m N ,
\end{equation}
which proves (\ref{ineq-reg2}). 

Part iii): For the sake of brevity, we write $\sol(x) := \sol_{v=0}(x)$. By \cite{Lieb+Yau1987} problem (\ref{def-EvN}), with $v=0$, has a minimizer that equals its symmetric-decreasing rearrangement, \ie $\sol(x) = \sol^*(x)$. In particular, $\sol(x)$ is a spherically symmetric, real-valued, nonincreasing function with $\sol(x) \geq 0$. It remains to show that $\sol(x) > 0$ holds. To see this, we put $\lambda = \mu$ in (\ref{eq-reg2}), which is possible by the proof of ii), and we obtain 
\begin{equation} \label{eq-reg3}
\sol = \big ( \Tm + \mu \big )^{-1} F(\sol) .
\end{equation}
Next we observe, by using functional calculus for the self-adjoint operator $\sqrt{-\Delta + m^2} : \Hone(\RR^3) \rightarrow \Lp{2}(\RR^3)$, that the following identity holds
\begin{equation} \label{eq-reg4}
\big ( \Tm + \mu \big )^{-1} = \int_0^\infty e^{-t \mu} e^{-t ( \sqrt{-\Delta+m^2} - m)} \diff t.
\end{equation}
By the explicit formula (\ref{eq-explicit}) for $v=0$, we see that the integral kernel, $e^{-t ( \sqrt{-\Delta+m^2} - m)}(x,y)$, is strictly positive. In view of (\ref{eq-reg3}), (\ref{eq-reg4}), and the fact that $F(\sol) \geq 0$, we conclude that $\sol(x) > 0$ holds for almost every $x \in \RR^3$. But since $\sol(x)$ is a nonincreasing and continuous function, we deduce that $\sol(x) > 0$ has to be true for all $x \in \RR^3$. This completes the proof of Theorem \ref{th-reg-decay}. \end{proof}

\section{Outlook}
\label{sec-outlook}

Our analysis presented so far serves as a basis for the upcoming work in \cite{FJL2005II} which  explores the effective motion of travelling solitary waves in an external potential. More precisely, we consider
\begin{equation} \label{eq-hartree-V}
i \partial_t \psi = \Tm \psi + V \psi - \Vconv{\psi} \quad \mbox{on $\RR^3$}.
\end{equation}
Here the external potential $V : \RR^3 \rightarrow \RR$ is assumed to be a smooth, bounded function with bounded derivatives. Note that its spatial variation introduces the length scale
\begin{equation}
\ell_{\mathrm{ext}} = \| \nabla V \|_\infty^{-1}   .
\end{equation}
In addition, another length scale, $\ell_{\mathrm{sol}}$, enters through the exponential decay of $\sol_v(x)$, \ie we have that
\begin{equation}
\ell_{\mathrm{sol}} = \delta^{-1},
\end{equation}
where $\delta > 0$ is the constant taken from Theorem \ref{th-reg-decay}. On intuitive grounds, one expects that if we have that
  \begin{equation} \label{eq-Lsol_small}
 \ell_{\mathrm{sol}} \ll \ell_{\mathrm{ext}} 
 \end{equation} 
holds, then solutions, $\psi(t,x)$, of (\ref{eq-hartree-V}) that are initially close to $\sol_v(x)$ should approximately behave like point-particles, at least on a large (but possibly finite) interval of time. 

We now briefly sketch how this heuristic picture of point-particle behavior of solitary waves is addressed by rigorous analysis in \cite{FJL2005II}. There we introduce a nondegeneracy assumption on the linearized operator 
\begin{equation}
L := \begin{pmatrix}
    L_1  &  0   \\
     0  & L_2 
\end{pmatrix} 
\end{equation}
acting on $\Lp{2}(\RR^3; \RR^2)$ with domain $\Sob{1}(\RR^3; \RR^2)$, where 
\begin{equation}
L_1 \xi := \Tmu \xi - \big ( \frac{1}{|x|} \ast \sol^2 \big ) \xi - 2  \big  ( \frac{1}{|x|} \ast (\sol \xi) \big ) \sol ,
\end{equation}
\begin{equation}
L_2 \xi := \Tmu \xi - \big ( \frac{1}{|x|} \ast \sol^2 \big ) \xi .
\end{equation}
Here $\sol(x) = \sol_{v=0}(x)$ is an unboosted ground state, which is chosen to be spherically symmetric and real-valued, by Theorem \ref{th-reg-decay}. The nondegeneracy condition is then
\begin{equation} \label{eq-kernel}
\mathrm{ker} (L) = \mathrm{span} \left \{  { 0 \choose \sol }, {\partial_{x_1} \sol \choose 0 }, { \partial_{x_2} \sol \choose 0}, { \partial_{x_3} \sol \choose 0 } \right \} .
\end{equation}
Under this assumption and for suitable external potentials of the form
\begin{equation}
V(x) := W(\epsilon x),
\end{equation}
we derive the following result in \cite{FJL2005II}: Let $\sol_{v_0,\mu_0}$ with $|v_0| \ll 1$ be given and choose $\epsilon \ll 1$ so that (\ref{eq-Lsol_small}) holds. Then for any initial datum, $\psi_0(x)$, such that
\begin{equation} \label{ineq-soleff}
||| \psi_0 - e^{i \vartheta_0} \sol_{v_0, \mu_0}(\cdot - a_0)  ||| \leq \epsilon, \quad \mbox{for some $\vartheta_0 \in \RR$ and $a_0 \in \RR^3$},
\end{equation}
where $||| \cdot |||$ is some weighted Sobolev norm, the corresponding solution, $\psi(t,x)$, of (\ref{eq-hartree-V}) can be uniquely written as
\begin{equation} \label{eq-eff} 
\psi(t,x) = e^{i \vartheta} [ \sol_{v,\mu}(x-a) + \xi(t,x-a) ], \quad \mbox{for $0 \leq t < C \epsilon^{-1}$}.
\end{equation} 
Here $||| \xi ||| = \mathcal{O}(\epsilon)$ holds and the time-dependent functions, $\{\vartheta, a, v, N \}$ with $N \equiv \mathcal{N}(\sol_{v,\mu})$, satisfy equations of the following form
\begin{equation} \label{eq-eff2}
\left \{ \begin{array}{ll} \dot{N} = \mathcal{O}(\epsilon^2), \quad & \dot{\vartheta} = \mu - V(a) + \mathcal{O}(\epsilon^2), \\
\dot{a} = v + \mathcal{O}(\epsilon^2), \quad & \gamma(\mu,v) \dot{v} = -\nabla V(a) + \mathcal{O}(\epsilon^2) .
\end{array} \right .
\end{equation}
The term $\gamma(\mu,v)$ can be viewed as an ``effective mass'' which takes relativistic effects into account.
 
Finally, we remark that the proof of (\ref{eq-eff}) and (\ref{eq-eff2}) makes extensive use of the Hamiltonian formulation of (\ref{eq-hartree-V}) and its associated symplectic structure restricted to the manifold of solitary waves. Moreover, assumption (\ref{eq-kernel}) enables us to derive additional properties of $\sol_v(x)$, for $|v| \ll 1$, such as cylindrical symmetry with respect to the $v$-axis, which is of crucial importance in the analysis presented in \cite{FJL2005II}.

\appendix
\section{Variational and Pseudo-Differential Calculus}
\label{sec-app-var}

In this section of the appendix, we collect and prove results needed for our variational and pseudo-differential calculus.

\subsection{Proof of Lemma \ref{lem-con-com}}

Let $(\psi_n)$ be a bounded sequence in $\Hhalf(\RR^3)$ with $\| \psi_n \|_2^2 = N$ for all $n$. Along the lines of \cite{Lions1984a}, we define the sequence, $(Q_n)$, of L\'evy concentration functions by
\begin{equation} \label{def-Levy}
Q_n(R) := \sup_{y \in \RR^3} \int_{|x-y| < R} | \psi_n |^2 \diff x, \quad \mbox{for $R \geq 0$.}
\end{equation}    
As stated in \cite{Lions1984a}, there exists a subsequence, $(Q_{n_k})$, such that 
\begin{equation} \label{eq-Qt}
\mbox{$Q_{n_k}(R) \rightarrow Q(R)$ as $ k \rightarrow \infty$ for all $R \geq 0$}, 
\end{equation}
where $Q(R)$ is a nonnegative, nondecreasing function. Clearly, we have that
\begin{equation}
\alpha := \lim_{R \rightarrow \infty} Q(R) \in [0,N].
\end{equation}
If $\alpha = 0$, then situation ii) of Lemma \ref{lem-con-com} arises as an direct consequence of definition (\ref{def-Levy}). If $\alpha = N$, then i) follows, see \cite{Lions1984a} for details.

Assume that $\alpha \in (0,N)$ holds, and let $\epsilon > 0$ be given. Suppose that $\xi, \phi \in \C{\infty}(\RR^3)$ with $0 \leq \phi, \xi \leq 1$ such that
\begin{equation}
\xi(x) \equiv 1 \quad \mbox{for $0 \leq |x| \leq 1$}, \qquad \xi(x) \equiv 0 \quad \mbox{for $|x| \geq 2$},
\end{equation}
\begin{equation}
\phi(x) \equiv 0 \quad \mbox{for $0 \leq |x| \leq 1$}, \qquad \phi(x) \equiv 1 \quad \mbox{for $|x| \geq 2$}.
\end{equation} 
Furthermore, we put $\xi_R(x) := \xi (x/R)$ and $\phi_R(x) := \phi (x/R)$, for $R > 0$, and we introduce
\begin{equation} \label{def-psi1-psi2}
\psi^1_k := \xi_{R_1}(\cdot - y_k) \psi_{n_k} \quad \mbox{and} \quad \psi^2_k := \phi_{R_k}(\cdot - y_k) \psi_{n_k}.
\end{equation}
As shown in \cite[Proof of Lemma III.1]{Lions1984a}, there exists 
\begin{equation}
R_1(\epsilon) \rightarrow \infty, \quad \mbox{as $\epsilon \rightarrow 0$},
\end{equation}
and a sequence, $(R_k)$, with 
\begin{equation}
R_k \rightarrow \infty, \quad \mbox{as $k \rightarrow \infty$},
\end{equation} 
such that $(\psi^1_k)$ and $(\psi^2_k)$ satisfy (\ref{ineq-con-com-mass}) and (\ref{ineq-con-com-supp}) in Lemma \ref{lem-con-com}. Moreover, we have that
\begin{equation} \label{ineq-cc1}
\int_{\RR^3} | \psi_{n_k} - (\psi^1_k + \psi^2_k) |^2 \diff x \leq 4 \epsilon ,
\end{equation}
for $k$ sufficiently large. 

By \cite[Theorem 7.16]{Lieb+LossII}, we see that $\psi^1_k$ and $\psi^2_k$ defined in (\ref{def-psi1-psi2}) are bounded in $\Hhalf(\RR^3)$. More precisely, using the technique of the proof given there and the explicit formula
\begin{equation}
\inner{f}{\T f} = \mathrm{(const.)}  \int_{\RR^3 \times \RR^3} \frac{ |f(x) - f(y)|^2}{|x-y|^4} \diff x \diff y , \quad \mbox{for $f \in \Hhalf(\RR^3)$,}
\end{equation}
we deduce that 
\begin{equation}
\nrmHhalf{gf} \leq C \big ( \|g\|_\infty + \| \nabla g \|_\infty \big ) \nrmHhalf{f} .
\end{equation}
Thus, we find that
\begin{equation}
\nrmHhalf{\psi^1_k} \leq C \big ( 1 + \frac{1}{R_1} \big ) \quad \mbox{and} \quad \nrmHhalf{\psi^2_k} \leq C \big ( 1 + \frac{1}{R_k} \big ),
\end{equation} 
for some constant $C = C(M)$, where $M = \sup_{k \geq 0} \nrmHhalf{\psi_{n_k}} < \infty$. Thus, $(\psi^1_{k})$ and $(\psi^2_k)$ are bounded sequences in $\Hhalf(\RR^3)$. This fact together with H\"older's and Sobolev's inequalities leads to
\begin{equation}
\big \| \psi_{n_k} - (\psi^1_k + \psi^2_k) \big \|_p \leq \delta_p(\epsilon), \quad \mbox{for $2 \leq p < 3$},
\end{equation}
where $\delta_p(\epsilon) \rightarrow 0$ as $\epsilon \rightarrow 0$. This proves (\ref{ineq-con-com-Lp}) in Lemma \ref{lem-con-com}.  

It remains to show property (\ref{ineq-con-com-liminf-T}) in Lemma \ref{lem-con-com}. Since
\begin{align} \label{ineq-liminf}
&\liminf_{k \rightarrow \infty} \left ( \inner{\psi_{n_k}}{(-m) \psi_{n_k}} - \inner{\psi^1_k}{(-m) \psi^1_k} - \inner{\psi^2}{(-m) \psi^2_k} \right ) \\
& \geq -mN +m (\alpha-\epsilon) + m(N-\alpha-\epsilon) \geq -2m \epsilon \rightarrow 0, \quad \mbox{as $\epsilon \rightarrow 0$}, 
\end{align}
we observe that it suffices to prove the claim
\begin{equation} \label{ineq-liminf-A}
\liminf_{k \rightarrow \infty} \left ( \inner{\psi_{n_k}}{A \psi_{n_k}} - \inner{\psi^1_k}{A \psi^1_k} - \inner{\psi^2_k}{A \psi^2_k} \right ) \geq -C(\epsilon),
\end{equation}
for some constant $C(\epsilon) \rightarrow 0$ as $\epsilon \rightarrow 0$, where
\begin{equation} \label{def-A}
A := \sqrt{-\Delta+m^2} + i (v \cdot \nabla) + \lambda,
\end{equation}
with $m \geq 0$, $v \in \RR^3$, $|v| < 1$, and $\lambda > 0$ is some constant so that 
\begin{equation}
A \geq (1-|v|) \T + \lambda \geq \lambda > 0.
\end{equation}
In view of (\ref{ineq-liminf}), adding any fixed $\lambda$ can be done without loss of generality.

Next, we recall definition (\ref{def-psi1-psi2}) and rewrite the left-hand side in (\ref{ineq-liminf-A}) as follows
\begin{equation} \label{ineq-liminf-A2}
\liminf_{k\rightarrow \infty} \innerb{\psi_{n_k}}{(A - \xi_k A \xi_k - \phi_k A \phi_k) \psi_{n_k}},
\end{equation}
where 
\begin{equation}
\xi_k(x) := \xi_{R_1}(x-y_k) \quad \mbox{and} \quad \phi_k(x) := \phi_{R_k}(x-y_k).
\end{equation}
Using commutators $[X,Y] := XY - YX$, we find that
\begin{align}
A - \xi_k A \xi_k - \phi_k A \phi_k & = A(1 - \xi_k^2 - \phi_k^2) - [\xi_k, A] \xi_k - [\phi_k, A] \phi_k \nonumber \\
& = \sqrt{A} (1-\xi^2_k -\phi^2_k) \sqrt{A} - \sqrt{A} [\sqrt{A},(\xi^2_k + \phi^2_k)] \nonumber \\
& \quad - [\xi_k, A] \xi_k - [\phi_k, A] \phi_k . \label{eq-A-commutator}
\end{align}
Note that $\sqrt{A} > 0$ holds, due to $A > 0$. By applying Lemma \ref{lem-commutator}, we obtain
\begin{equation}
\big \| [ \xi_k, A ] \big \|_{\Lp{2} \rightarrow \Lp{2}} \leq C \| \nabla \xi_k \|_\infty \leq \frac{C}{R_1},
\end{equation}
\begin{equation}
\big \| [ \phi_k, A ] \big \|_{\Lp{2} \rightarrow \Lp{2}} \leq C \| \nabla \phi_k \|_\infty \leq \frac{C}{R_k},
\end{equation}
To estimate the remaining commutator in (\ref{eq-A-commutator}), we use (\ref{eq-fro}) in the proof of Lemma \ref{lem-commutator} to find that
\begin{align}
\big \| [ \sqrt{A}, (\xi^2_k + \phi^2_k) ] \big \|_{\Lp{2} \rightarrow \Lp{2}} & \leq C \left( \frac{1}{R_1} + \frac{1}{R_k} \right ) \int_0^\infty \sqrt{s} \big \| \frac{1}{(s+A)} \big \|^2_{\Lp{2} \rightarrow \Lp{2}} \diff s \\
& \leq  C \left( \frac{1}{R_1} + \frac{1}{R_k} \right ) \int_0^\infty \frac{\sqrt{s}}{(s+\lambda)^2} \diff s \\
& \leq C \left( \frac{1}{R_1} + \frac{1}{R_k} \right )
\end{align}
Returning to (\ref{ineq-liminf-A2}) and using that $\nrmHhalf{\psi_{n_k}} \leq C$, we conclude, for $k$ large, that
\begin{align}
\inner{\psi_{n_k}}{(A - \xi_k A \xi_k - \phi_k A \phi_k) \psi_{n_k}} & \geq  \inner{\sqrt{A}\psi_{n_k}}{(1-\xi^2_k-\phi^2_k) \sqrt{A}\psi_{n_k}} - C \left (\frac{1}{R_1} + \frac{1}{R_k} \right ) \\
& \geq - C \left (\frac{1}{R_1} + \frac{1}{R_k} \right ), 
\end{align}
since $(1-\xi^2_k-\phi_k^2)(x) \geq 0$ when $k$ is sufficiently large. Finally, we note that $R_k \rightarrow \infty$ as $k \rightarrow \infty$ as well as $R_1(\epsilon) \rightarrow \infty$ as $\epsilon \rightarrow 0$ holds, which leads to
\begin{equation}
\liminf_{k \rightarrow \infty} \innerb{\psi_{n_k}}{(A - \xi_k A \xi_k - \phi_k A \phi_k) \psi_{n_k}} \geq -C(\epsilon) \rightarrow 0, \quad \mbox{as $\epsilon \rightarrow 0$.}
\end{equation} 
The proof of Lemma \ref{lem-con-com} is now complete. \hfill $\qed$

\subsection{Technical Details for the Proof of Theorem \ref{th-existence}}

\begin{lemma} \label{lem-vanish}
Let $(\psi_n)$ satisfy the assumptions of Lemma \ref{lem-con-com}. Furthermore, suppose that there exists a subsequence, $(\psi_{n_k})$, that satisfies part ii) of Lemma \ref{lem-con-com}. Then
\[
\lim_{k \rightarrow \infty} \Vpot{\psi_{n_k}} = 0 .
\]
\end{lemma}

\medskip
\noindent
{\bf Remark.} A similar statement can be found in \cite{Lions1984a} in the context of other variational problems. For the sake of completeness, we present its proof for the situation at hand.

\begin{proof}[Proof of Lemma \ref{lem-vanish}]
Let $(\psi_{n_k})$ be bounded sequence in $\Hhalf(\RR^3)$ such that
\begin{equation}
\int_{\RR^3} |Ê\psi_{n_k}|^2 \diff x = N, \quad \mbox{for all $k \geq 0$},
\end{equation}
and assume that $(\psi_{n_k})$ satisfies part ii) in Lemma \ref{lem-con-com}, \ie
\begin{equation}
\lim_{k \rightarrow \infty} \sup_{y \in \RR^3} \int_{|x-y| < R} |\psi_{n_k}|^2 \diff x = 0, \quad \mbox{for all $R > 0$}.  
\end{equation}
For simplicity, let $\psi_k := \psi_{n_k}$.

We introduce  
\begin{equation}
f_\delta(x) := |x|^{-1} \chi(x)_{\{ |x|^{-1} \geq \delta \}}, \quad \mbox{with $\delta > 0$},
\end{equation}
where $\chi_A$ denotes the characteristic function of the set $A \subset \RR^3$. This definition leads to
\begin{equation} \label{ineq-vanish1}
\Vpot{\psi_k} \leq \delta C + \int_{\RR^3 \times \RR^3} |\psi_k(x)|^2 |\psi_k(y)|^2 f_\delta(x-y) \diff x \diff y,
\end{equation}
where $C$ is some constant. For $R > 0$ and $\delta > 0$, let
\begin{equation}
g_\delta^R(x) := \min \{ f_\delta(x), R \}, 
\end{equation}
\begin{equation}
 f^R_\delta(x) := \max \{ f_\delta(x) - R, 0 \} \chi(x)_{\{ |x| \leq R \}} + f_\delta(x) \chi(x)_{\{ |x| > R \} }.
\end{equation}
Notice that $f_\delta \leq g_\delta^R \chi_{\{ |x| \leq R \}} + f^R_\delta$ holds. In view of (\ref{ineq-vanish1}), this leads to
\begin{align}
\Vpot{\psi_k} & \leq \delta C + \int_{\RR^3 \times \RR^3} | \psi_k(x) |^2 | \psi_k(y) |^2 g_\delta^R(x-y) \chi(x-y)_{\{ |x-y| \leq R \}} \diff x \diff y \nonumber \\
& \quad + \| \psi_k \|^4_{8/3} \| f^R_\delta \|_2 \nonumber \\
& =: \delta C + I + II,
\end{align}
using Young's inequality and that $f^R_\delta \in \Lp{2}(\RR^3)$. By our assumption on $(\psi_k)$, we find that
\begin{align*}
I \leq R \int_{\RR^3} |\psi_k(x)|^2 \diff x \int_{|x-y| \leq R} | \psi_k(y) |^2 \diff y \rightarrow 0, \quad \mbox{as $k \rightarrow \infty$} .
\end{align*} 
Furthermore, we have that
\begin{equation}
II \leq C \| f^R_\delta \|_2,
\end{equation}
by Sobolev's inequalities and the fact that $(\psi_k)$ is bounded in $\Hhalf(\RR^3)$.  Thus, we obtain
\begin{equation}
0 \leq \Vpot{\psi_k} \leq \delta C  + C \| f^R_\delta \|_{2} + r(k), \quad \mbox{for all $\delta, R > 0$},
\end{equation}
where $r(k) \rightarrow 0$ as $k \rightarrow \infty$. Since $\big \| f^R_\delta Ê\big \|_{2} \rightarrow 0$ as $R \rightarrow \infty$, for each fixed $\delta > 0$, the assertion of Lemma \ref{lem-vanish} follows by letting $R \rightarrow \infty$ and then sending $\delta$ to 0.\end{proof}

\begin{lemma} \label{lem-dicho}
Suppose that $\epsilon > 0$. Let $(\psi_n)$ satisfy the assumptions of Lemma \ref{lem-con-com} and let $(\psi_{n_k})$ be a subsequence that satisfies part iii) with sequences $(\psi^1_k)$ and $(\psi^2_k)$. Then, for $k$ sufficiently large, 
\begin{align*}
- \Vpot{\psi_{n_k}} & \geq - \Vpot{\psi^1_k} \nonumber \\
& \quad - \Vpot{\psi^2_k} - r_1(k) - r_2(\epsilon) , 
\end{align*}
where $r_1(k) \rightarrow 0$ as $k \rightarrow \infty$ and $r_2(\epsilon) \rightarrow 0$ as $\epsilon \rightarrow 0$.
\end{lemma}

\begin{proof}[Proof of Lemma \ref{lem-dicho}]
Let $\epsilon > 0$ and suppose that $(\psi_{n_k})$, $(\psi^1_k)$, and $(\psi^2_k)$ satisfy the assumptions stated above. Introducing
\begin{equation}
\beta_{k} := \psi_{n_k} - ( \psi^1_k + \psi^2_k )
\end{equation}
and expanding the squares, we find that 
\begin{align}
\Vpot{\psi_{n_k}} & = \Vpot{\psi^1_k} \nonumber \\
& \quad + \Vpot{\psi^2_k}  + \sum_{n=0}^4 I_n,
\end{align}
where
\begin{align}
I_0 & = 2 \int_{\RR^3} \big ( \frac{1}{|x|} \ast |\psi^1_k|^2 \big ) | \psi^2_k |^2 \diff x + 4 \int_{\RR^3} \big ( \frac{1}{|x|} \ast (\RE \bar{\psi}^1_k \psi^2_k) \big ) (\RE \bar{\psi}^1_k \psi^2_k ) \diff x\\
 & \quad + 4 \int_{\RR^3} \big ( \frac{1}{|x|} \ast |\psi^1_k|^2 \big ) ( \RE \bar{\psi^1_k} \psi^2_k ) \diff x  +  4\int_{\RR^3} \big ( \frac{1}{|x|} \ast |\psi^2_k|^2 \big ) ( \RE \bar{\psi^1_k} \psi^2_k ) \diff x, \label{def-I}  \\
I_1 & = 4 \int_{\RR^3} \big ( \frac{1}{|x|} \ast |\psi^1_k + \psi^2_k|^2 )  (\RE \bar{\beta}_k (\psi^1_k + \psi^2_k)) \diff x, \label{def-II} \\
I_2 & = 4 \int_{\RR^3} \big ( \frac{1}{|x|} \ast (\RE  \bar{\beta}_k (\psi^1_k + \psi^2_k)) \big ) (\RE \bar{\beta}_k(\psi^1_k + \psi^2_k)) \diff x \nonumber \\
& \quad + 2 \int_{\RR^3} \big ( \frac{1}{|x|} \ast | \psi^1_k + \psi^2_k |^2 \big ) |\beta_k|^2 \diff x, \label{def-III}  \\
I_3 &= 4 \int_{\RR^3} \big ( \frac{1}{|x|} \ast |\beta_k|^2 \big ) (\RE \bar{\beta}_k (\psi^1_k + \psi^2_k)) \diff x , \label{def-IV} \\
I_4 & = \int_{\RR^3} \big ( \frac{1}{|x|} \ast |\beta_k|^2 \big ) |\beta_k|^2 \diff x \label{def-V} .
\end{align}

To estimate $I_0$, we notice that if $k$ is sufficiently large then $\psi^1_k$ and $\psi^2_k$ have disjoint supports receding from each other, \ie
\begin{equation} \label{eq-dis-supp}
d_k := \mathrm{dist} \, ( \mathrm{supp} \, \psi^1_k, \mathrm{supp} \, \psi^2_k ) \rightarrow \infty, \quad \mbox{as $ k \rightarrow \infty$};
\end{equation}
see the proof of Lemma \ref{lem-con-com} in Sect.~A.1. Thus, the last three terms of the right-hand side in (\ref{def-I}) equal 0 if $k$ is large, since $\bar{\psi}^1_k \psi^2_k = 0$ a.\,e.~if $k$ is sufficiently large. Also by (\ref{eq-dis-supp}), we infer
\begin{align}
\Big | \int_{\RR^3 \times \RR^3} | \psi^1_k(x)|^2 \frac{1}{|x-y|} |\psi^2_k(y)|^2 \diff x \diff y \Big | & = \Big |  \int_{\RR^3 \times \RR^3}  | \psi^1_k(x)|^2 \frac{\chi(x-y)_{\{ |x-y| \geq d_k\}} }{|x-y|} |\psi^2_k(y)|^2 \diff x \diff y \Big | \nonumber \\
& \leq \big \| \psi^1_k \big \|_2^2 \big \| \psi^2_k \big \|_2^2 \big \| |x|^{-1} \chi(x)_{\{ |x| \geq d_k\} } \big \|_\infty \nonumber \\
& \leq \frac{C}{d_k} \rightarrow 0, \quad \mbox{as $k \rightarrow \infty$}, 
\end{align} 
using Young's inequality. Thus we have shown that 
\begin{equation}
|I_0| \leq r_1(k) \rightarrow 0, \quad \mbox{as $k \rightarrow \infty$}.
\end{equation}

The remaining terms $I_1$--$I_4$ can be controlled by the Hardy-Littlewood-Sobolev inequality and H\"older's inequality as follows
\begin{align}
|I_1| & \leq C ( \| \psi^1_k \|_{12/5}^3 + \| \psi^2_k \|_{12/5}^3 ) \| \beta_k \|_{12/5}, \qquad |I_2| \leq C ( \|\psi^1_k \|_{12/5}^2 + \| \psi_k^2 \|^2_{12/5} ) \| \beta_k \|_{12/5}^2 , \\
|I_3| & \leq C ( \| \psi^1_k \|_{12/5} + \| \psi^2_k \|_{12/5} ) \| \beta_k \|_{12/5}^3, \qquad |I_4| \leq C \| \beta \|_{12/5}^4 .
\end{align}
We notice that $\| \psi^1_k \|_{12/5}$ and $\| \psi^2_k \|_{12/5}$ are uniformly bounded, by Sobolev's inequality and the $\Hhalf$-boundedness of these sequences. Furthermore, we have that
\begin{equation}
\| \beta_k \|_{12/5} \leq r_2(\epsilon) \rightarrow 0, \quad \mbox{as $\epsilon \rightarrow 0$},
\end{equation}
by part iii) of Lemma \ref{lem-con-com}. Hence we conclude that
\begin{equation}
|I_1 + \cdots + I_4| \leq r_2(\epsilon) \rightarrow 0, \quad \mbox{as $\epsilon \rightarrow 0$},
\end{equation}
which proves Lemma \ref{lem-dicho}. \end{proof}

\subsection{Commutator Estimate}

An almost identical result is needed in \cite{Froehlich+Lenzmann2005}, but we provide its proof again. 

\begin{lemma} \label{lem-commutator}
Let $m \geq 0$, $\vel \in \RR^3$, and define $A_v := \sqrt{-\Delta + m^2} + i (v \cdot \nabla)$. Furthermore, suppose that $f(x)$ is a locally integrable and that its distributional gradient, $\nabla f$, is an $\Lp{\infty}(\RR^3)$ vector-valued function. Then we have that
\[
 \| [A_v, f]  \|_{\Lp{2} \rightarrow \Lp{2}} \leq C_v \| \nabla f \|_\infty,
\]
for some constant $C_v$ that only depends on $v$.
\end{lemma}

\medskip
\noindent
{\bf Remark.} This result can be deduced by means of Calder\'on--Zygmund theory for singular integral operators and its consequences for pseudo-differential operators (see, \eg \cite[Section VII.3]{Stein1993}). We give an elementary proof which makes good use of the spectral theorem, enabling us to write the commutator in a convenient way. 

\begin{proof}
Since $[i( v \cdot \nabla), f] = i v \cdot \nabla f$ holds, we have that 
\begin{equation}
\| [i (v \cdot \nabla) , f] \|_{\Lp{2} \rightarrow \Lp{2}} \leq |v| \| \nabla f \|_\infty.
\end{equation}
Thus, it suffices to prove our assertion for $A := A_{v=0}$, \ie
\begin{equation}
A := \sqrt{p^2 + m^2} , \quad \mbox{where $p = - i \nabla$} .
\end{equation}

Since $A$ is a self-adjoint operator on $\Lp{2}(\RR^3)$ (with domain $\Sob{1}(\RR^3)$), functional calculus (for measurable functions) yields the formula
\begin{equation}
A^{-1} = \frac{1}{\pi} \int_0^\infty \frac{1}{\sqrt{s}} \frac{\diff s}{A^2+s} .
\end{equation}
Due to this fact and $A=A^2 A^{-1}$, we obtain the formula
\begin{equation} \label{eq-fro}
[ A, f ] = \frac{1}{\pi} \int_0^\infty \frac{\sqrt{s}}{A^2 + s} [A^2, f] \frac{\diff s}{A^2 + s} .
\end{equation} 
Clearly, we have that $[A^2, f] = [p^2,f] = p \cdot [p,f] + [p,f] \cdot p$, which leads to
\begin{equation} 
[A,f] = \frac{1}{\pi} \int_0^\infty \frac{\sqrt{s}}{p^2 + m^2 + s} \big ( p \cdot [p,f] + [p,f] \cdot p \big ) \frac{\diff s}{p^2 + m^2 + s} .
\end{equation}
Moreover, since $[p,f] = -i \nabla f$ holds, we have that
\begin{equation}
\big \| \big [ \frac{1}{p^2+m^2+s}, [p,f] \big ] \big \|_{L^{2} \rightarrow L^2}  \leq \frac{2}{s} \| \nabla f\|_\infty .
\end{equation}
Hence we find, for arbitrary Schwartz functions $\xi, \eta \in \mathcal{S}(\RR^3)$, that
\begin{align}
& \Big | \innerb{\xi} {\int_0^\infty \frac{\sqrt{s}}{p^2 + m^2 + s} \big ( [p,f] \cdot p \big ) \frac{\diff s}{p^2 + m^2 + s} \, \eta } \Big | \label{ineq-L2bound} \\
 & \leq \Big | \innerb{[p,f] \xi}{p \int_0^\infty \frac{\sqrt{s} \diff s}{(p^2+m^2+s)^2} \, \eta} \Big | + \Big | \innerb{\xi}{ \int_0^\infty \big [ \frac{1}{p^2+m^2+s}, [p,f] \big ] \cdot \frac{  p \sqrt{s} \diff s}{p^2+m^2+s} \, \eta } \Big | \nonumber \\
 & \leq \big \| [p,f] \xi \big \|_2 \big \|  \int_0^\infty \frac{p \sqrt{s}  \diff s}{(p^2+m^2+s)^2}  \, \eta \big \|_2   + 2 \| \xi \|_2 \| \nabla f \|_\infty \big \|  \int_0^\infty \frac{ p  \diff s}{\sqrt{s} (p^2+m^2+s)} \, \eta \big \|_2. \nonumber
\end{align}
Evaluation of the $s$-integrals yields
\begin{align}
\mbox{(\ref{ineq-L2bound})} & \leq C \big \| \nabla f\|_\infty \| \xi \big \|_2  \big \| \frac{  p }{\sqrt{p^2+m^2}} \eta \big \|_2   \leq C \| \nabla f \|_\infty  \| \xi \|_2 \| \eta \|_2 .
\end{align}
The same estimate holds if $[p,f] \cdot p$ is replaced by $p \cdot [p,f]$ in (\ref{ineq-L2bound}). Thus, we have found that
\begin{equation}
\big | \inner{\xi}{ [A,f] \eta}  \big | \leq C \| \nabla f \|_\infty \| \xi \|_2 \| \eta \|_2, \quad \mbox{for $\xi, \eta \in \mathcal{S}(\RR^3)$},
\end{equation}
with some constant $C$ independent of $m$. Since $\mathcal{S}(\RR^3)$ is dense in $L^2(\RR^3)$, the assertion for the $L^2$-boundedness of $[A,f]$ now follows. This completes the proof of Lemma \ref{lem-commutator}.\end{proof}

\subsection{Lower Semicontinuity}

\begin{lemma} \label{lem-wlsc}
Suppose that $m > 0$, $\vel \in \RR^3$, with $|\vel| < 1$. Then the functional
\[
\mathcal{T}(\psi) := \innerb{\psi}{ \big ( \sqrt{-\Delta + m^2} \big ) \psi} + \innerb{\psi}{ i (v \cdot \nabla) \psi}
\]
is weakly lower semicontinuous on $\Hhalf(\RR^3)$, \ie if $\psi_k \rightharpoonup \psi$ weakly in $\Hhalf(\RR^3)$ as $k \rightarrow \infty$, then
\[
\liminf_{k \rightarrow \infty} \mathcal{T}(\psi_k) \geq \mathcal{T}(\psi) .
\]
Moreover, if $\lim_{k \rightarrow \infty} \mathcal{T}(\psi_k) = \mathcal{T}(\psi)$ holds, then $\psi^k \rightarrow \psi$ strongly in $\Hhalf(\RR^3)$ as $k \rightarrow \infty$.
\end{lemma}

\begin{proof}[Proof of Lemma \ref{lem-wlsc}]
Assume that $m > 0$, $\vel \in \RR^3$, with $|\vel| < 1$ holds. By Fourier transform and Plancherel's theorem, we have that
\begin{equation} \label{eq-Tplancherel}
\mathcal{T}(\psi) =  \int_{\RR^3} | \widehat{\psi}(k) |^2 \left ( \sqrt{k^2 + m^2} - (v \cdot k) \right ) \diff k.
\end{equation}
We notice that
\begin{equation} \label{ineq-wlsc}
c_1 ( |k| + m ) \leq \sqrt{k^2 + m^2} - (v \cdot k) \leq c_2 (|k| + m) ,
\end{equation}
for some suitable constants $c_1, c_2 > 0$, where the lower bound follows from the inequality $\sqrt{k^2 + m^2} \geq (1-\delta) |k| + \delta m$, with $0 < \delta < 1$, and the fact that $|v| < 1$ holds. Thus, 
\begin{equation}
\| \psi \|_\mathcal{T} := \sqrt{\mathcal{T}(\psi)}
\end{equation}
defines a norm that is equivalent to $\nrmHhalf{\cdot}$. Consequently, the notion of weak and strong convergence for these norms coincide. Finally, by (\ref{eq-Tplancherel}), we identify $\| \psi \|_\mathcal{T}$ with the $L^2$-norm of $\widehat{\psi}$ taken with respect to the integration measure
\begin{equation}
\diff \mu = \left ( \sqrt{k^2 + m^2} - (v \cdot k) \right ) \diff k.
\end{equation} 
The assertion of Lemma \ref{lem-wlsc} now follows from corresponding properties of the $\Lp{2}(\RR^3, \mu)$-norm; see, \eg \cite[Theorem 2.11]{Lieb+LossII} for $\Lp{p}(\Omega, \mu)$-norms, where $\Omega$ is a measure space with positive measure, $\mu$, and $1 < p < \infty$.   \end{proof}

\section{Best Constant and Optimizers for Inequality (\ref{ineq-Sopt})} 
\label{sec-app-best}

\begin{lemma} \label{lem-constant}
For any $v \in \RR^3$ with $|v| < 1$, there exists an optimal constant, $S_v$, such that 
\begin{equation} \label{ineq-constant}
\Vpot{\psi} \leq S_v \innerb{\psi}{\big ( \T + i v \cdot \nabla \big ) \psi} \innerb{\psi}{\psi}
\end{equation}
holds for all $\psi \in \Hhalf(\RR^3)$. Moreover, we have that
\begin{equation} \label{eq-Sv}
S_v = \frac{2}{ \inner{Q_v}{Q_v} } ,
\end{equation} 
where $Q_v \in \Hhalf(\RR^3)$, $Q_v \not \equiv 0$, is an optimizer for (\ref{ineq-constant}) and it satisfies 
\begin{equation} \label{eq-EL-Qv}
\T \, Q_v + i (v \cdot \nabla) Q_v - \big ( \frac{1}{|x|} \ast |Q_v|^2 \big ) Q_v = - Q_v .
\end{equation}
In addition, the following estimates hold:
\begin{equation}
S_{v=0} < \frac{\pi}{2} \quad \mbox{and} \quad S_{v=0} \leq S_v \leq (1-|v|)^{-1} S_{v=0}. 
\end{equation}
\end{lemma} 

\begin{proof}[Proof of Lemma \ref{lem-constant}]
Let $v \in \RR^3$ with $|v| < 1$ be fixed and consider the unconstrained minimization problem
\begin{equation} \label{def-Kv}
\frac{1}{S_v} := \inf_{\psi \in \Hhalf(\RR^3), \psi \not \equiv 0} \frac{ \inner{\psi}{(\T + i v \cdot \nabla) \psi} \inner{\psi}{\psi}}{\int_{\RR^3} (|x|^{-1} \ast |\psi|^2) |\psi|^2 \diff x} .  
\end{equation}
For $v=0$, a variational problem equivalent to (\ref{def-Kv}) is studied in \cite[Appendix B]{Lieb+Yau1987} by using strict rearrangement inequalities that allow restriction to radial functions. For $v \neq 0$, we have to depart this line of argumentation and we employ (similarly to the discussion of (\ref{def-EvN}) in Sect.~\ref{sec-existence}) concentration-compactness-type methods. 

By scaling properties of (\ref{def-Kv}), it suffices to prove the existence of a minimizer with $\inner{\psi}{(\T + i v \cdot \nabla) \psi}$ and $\inner{\psi}{\psi}$ fixed. Thus, we introduce the constrained minimization problem, which is equivalent to (\ref{def-Kv}), as follows     
 \begin{equation} \label{def-Iv}
I_v(\alpha, \beta) := \inf \big \{ -\Vpot{\psi} : \inner{\psi}{\psi} = \alpha, \; \inner{\psi}{(\T + i v \cdot \nabla) \psi} = \beta \big \} ,
\end{equation}
where $\alpha > 0$ and $\beta > 0$. In particular, it is sufficient to show that $I_v(\alpha=1, \beta=1)$ is finite and attained so that
\begin{equation}
S_v = - I_v(1,1) .
\end{equation}  

In fact, we will show that all minimizing sequences for $I(1,1)$ are relatively compact in $\Hhalf(\RR^3)$ up to translations. In turn, this relative compactness implies that all minimizing sequences for problem (\ref{def-Kv}) are relatively compact in $\Hhalf(\RR^3)$ up to translations and rescalings: For any minimizing sequence, $(\psi_n)$, for (\ref{def-Kv}), there exist sequences, $\{(y_k), (a_k), (b_k) \}$, with $y_k \in \RR^3, 0 \neq a_k \in \mathbb{C}, 0 \neq b_k \in \RR$, such that
\begin{equation}
a_k \psi_{n_k} \big ( b_k( \cdot+ y_k) \big ) \rightarrow Q_v \quad \mbox{strongly in $\Hhalf(\RR^3)$ as $k \rightarrow \infty$},
\end{equation} 
along a suitable subsequence, $(\psi_{n_k})$, and $Q_v$ minimizes (\ref{def-Kv}).  

First we show that $I(\alpha, \beta)$ is indeed finite. The Hardy--Littlewood--Sobolev inequality (see, \eg \cite{Lieb+LossII}) implies
\begin{equation}
\Vpot{\psi} \leq C \| |\psi|^2 \|_{6/5}^2 = C \| \psi \|_{12/5}^4 \leq C \inner{\psi}{\T \psi} \inner{\psi}{\psi} 
\end{equation}
where we use Sobolev's inequality $\| \psi \|_3^2 \leq C \inner{\psi}{\T \, \psi}$ in $\RR^3$ and H\"older's inequality. Since $\inner{\psi}{\T \psi} \leq (1-|v|)^{-1} \inner{\psi}{(\T + i v \cdot \nabla) \psi}$, we deduce that
\begin{equation}
I(\alpha, \beta) \geq -C \alpha \beta > - \infty,
\end{equation} 
for some constant $C$. On the other hand, we have that
\begin{equation} \label{ineq-Ineg}
I(\alpha, \beta) < 0,
\end{equation} 
since $\int_{\RR^3} ( |x|^{-1} \ast |\psi|^2 ) |\psi|^2 \diff x \neq 0$ when $\psi \not \equiv 0$.

Next, we show that $I_v(1,1)$ is attained. Let $(\psi_n)$ be a minimizing sequence for $I_v(1,1)$. In order to invoke Lemma \ref{lem-con-com}, we notice that $\int_{\RR^3} |\psi_n|^2 \diff x = 1$ and that $(\psi_n)$ is bounded in $\Hhalf(\RR^3)$, since $\inner{\psi}{(\T + i v \cdot \nabla) \psi}$ is equivalent to $\inner{\psi}{\T \psi}$ when $|v| < 1$, by (\ref{ineq-wlsc}) with $m=0$.

Let us suppose now that case ii) of Lemma \ref{lem-con-com} occurs. Referring to Lemma \ref{lem-vanish}, we conclude that $I(1,1) = 0$ holds, which contradicts (\ref{ineq-Ineg}) . Next, let us assume that dichotomy occurs for a subsequence of $(\psi_n)$, \ie property iii) of Lemma \ref{lem-con-com} holds. Using Lemma \ref{lem-dicho} and the $\liminf$-estimate stated in iii) of Lemma \ref{lem-con-com} and by taking the limit $\epsilon \rightarrow 0$, we conclude that
\begin{equation} \label{eq-Iv-wrong}
I_v(1,1) \geq I_v(\alpha, \beta) + I_v(1-\alpha, 1-\beta),
\end{equation}
for some $\alpha \in (0,1)$ and $\beta \in [0,1]$. On the other hand, we have the scaling behaviour
\begin{equation} \label{eq-Iv-true}
I_v(\alpha, \beta) = \alpha \beta I_v(1,1) < 0,
\end{equation}
which follows from (\ref{def-Iv}) and rescaling $\psi(x) \mapsto a \psi(bx)$ with $a,b > 0$. Combining (\ref{eq-Iv-wrong}) with (\ref{eq-Iv-true}) we get a contradiction. Therefore dichotomy for minimizing sequences is ruled out. 

In summary, we see that any minimizing sequence, $(\psi_n)$, for $I_v(1,1)$ contains a subsequence, $(\psi_{n_k})$, with a sequence of translations, $(y_k)$, satisfying property i) of Lemma \ref{lem-con-com}. Similarly to the proof of Theorem \ref{th-existence}, we conclude that $\psi_{n_k}(\cdot + y_k) \rightarrow \widetilde{Q}_v$ strongly in $\Hhalf(\RR^3)$ as $k \rightarrow \infty$, where $\widetilde{Q}_v \in \Hhalf(\RR^3)$ is a minimizer for $I_v(1,1)$.   

To show that the best constant, $S_v$, is given by (\ref{eq-Sv}) with $Q_v$ minimizing (\ref{def-Kv}) and satisfying (\ref{eq-EL-Qv}), let us denote the minimizer constructed above for $I(1,1)$ by $\widetilde{Q}_v$. Since $\widetilde{Q}_v$ also minimizes the unconstrained problem (\ref{def-Kv}), it has to satisfy the corresponding Euler-Lagrange equation which reads as follows
\begin{equation}
\T \, \widetilde{Q}_v + i v \cdot \nabla \widetilde{Q}_v - \frac{2}{ S_v} \big ( \frac{1}{|x|} \ast |\widetilde{Q}_v|^2 \big ) \widetilde{Q}_v + \widetilde{Q}_v = 0,
\end{equation} 
where we use that $\inner{\widetilde{Q}_v}{(\T + i v \cdot \nabla ) \widetilde{Q}_v} = 1$ and $\inner{\widetilde{Q}_v}{\widetilde{Q}_v} = 1$ holds. By putting $Q_v = \sqrt{2} S_v^{-1/2} \widetilde{Q}_v$, we see that $Q_v$ minimizes (\ref{def-Kv}) and satisfies (\ref{eq-Sv}). Moreover, we have that $\inner{Q_v}{Q_v} = 2 / S_v$ holds. 

Finally, we turn to the estimates for $S_v$ stated in Lemma \ref{lem-constant}. That $S_{v=0} < \pi/2$ holds follows from the appendices in \cite{Lieb+Yau1987, Lenzmann2005LWP}. To see that $S_v \leq (1-|v|)^{-1} S_{v=0}$ is true, we use the estimate $\T \leq (1-|v|)^{-1} (\T + i v \cdot \nabla)$. Moreover, it is known from the discussion in \cite{Lenzmann2005LWP} that if $v=0$ the minimizer, $Q_{v=0}$, for (\ref{def-Ev}) can be chosen to be radial (by symmetric rearrangement). This implies that $\inner{Q_{v=0}}{\nabla Q_{v=0}} = 0$, which leads to $S_{v=0} \leq S_v$. \end{proof}

\section{Exponential Decay}
\label{sec-app-exp}

In this section, we address pointwise exponential decay for solutions, $\sol \in \Hhalf(\RR^3)$, of the nonlinear equation
\begin{equation} \label{eq-exp}
\Tm \sol + i (v \cdot \nabla) \sol - \big ( \frac{1}{|x|} \ast |\sol|^2 \big ) \sol = - \mu \sol.
\end{equation} 
Clearly, $\sol(x)$ is an eigenfunction for the Schr\"odinger type operator
\begin{equation}
H = H_0 + V,
\end{equation}
where 
\begin{equation} \label{def-H0}
H_0:= \Tm + i (v \cdot \nabla) \quad \mbox{and} \quad V := - \big ( \frac{1}{|x|} \ast |\sol|^2 ) .
\end{equation}
By using the bootstrap argument for regularity (presented in the proof of Theorem \ref{th-reg-decay})), we have that $\sol \in \Sob{s}(\RR^3)$ for all $s \geq 1/2$, which shows in particular that $\sol$ is smooth. Investigating the spectrum of $H_0$ we find that
\begin{equation}
\sigma ( H_0 ) = \sigma_{\mathrm{ess}} ( H_{0} ) = [ \Sigma_v, \infty),
\end{equation}
where the bottom of the spectrum is given by
\begin{equation}
\Sigma_v =(\sqrt{1-v^2} - 1) m .
\end{equation}
To see this, we remark that the function 
\begin{equation}
f(k) = (k^2 + m^2)^{1/2} -m - v\cdot k
\end{equation}
obeys $f(k) \geq  (\sqrt{1-v^2}) m - m$ with equality for $k = (mv/\sqrt{1-v^2}) \hat{v}$, where $\hat{v} = v/|v|$.  

We have the following result.
\begin{lemma} \label{lem-exp-decay}
Suppose that $m > 0$, $\vel \in \RR^3$, and $|\vel| < 1$. Furthermore, let $\sol \in \Hhalf$ be a solution of (\ref{eq-exp}) with $-\mu < \Sigma_v$. Then, for every $0 < \delta < \min \big \{ m, \frac{\Sigma_v + \mu}{\sqrt{1-v^2}} \big \}$, there exists $0 < C(\delta) < \infty$ such that
\[
| \sol(x) | \leq C e^{-\delta |x|}
\]
holds for all $x \in \RR^3$.
\end{lemma}

\begin{proof}
We rewrite (\ref{eq-exp}) as follows
\begin{equation} \label{eq-exp2}
 \sol = -(H_{0} + \mu)^{-1} V \sol,
\end{equation}
where $H_{0}$ and $V$ are defined in (\ref{def-H0}). Note that $(H_{0}+ \mu)^{-1}$ exists, since we have that $\mu \not \in \sigma(H_{0})$ holds, by the assumption that $-\mu < \Sigma_v$. We consider the Green's function, $G_\mu(x-y)$, given by 
\begin{equation}
G_\mu(x-y) = \mathcal{F}^{-1} \Big [ \frac{1}{\sqrt{k^2 +m^2} - m - v \cdot k +\mu} \Big ](x-y), 
\end{equation}
where $\mathcal{F} : \mathcal{S}' \rightarrow \mathcal{S}'$ denotes the Fourier transform. Since the function $1/\sqrt{k^2+m^2}\ldots$ does not belong to $\Lp{1}(\RR^3)$, we cannot use Payley--Wiener type theorems directly to deduce pointwise exponential decay for $G_\mu(z)$ in $|z|$. To overcome this difficulty, we first notice that 
\begin{equation} \label{eq-H00}
(H_{0} + \mu)^{-1} = \int_0^\infty e^{-t\mu} e^{-t H_{0}} \diff t = \int_0^\infty e^{-t(\mu-m)} e^{-t(\sqrt{p^2 + m^2} - v \cdot p)} \diff t,
\end{equation} 
by self-adjointness of $H_{0}$ and functional calculus. Here and in what follows, we put $p = -i \nabla$ for convenience. By using the explicit formula for the Fourier transform of $\exp \{-t \sqrt{k^2+m^2} \}$ (see \eg \cite{Lieb+LossII}) in $\RR^3$ and by analytic continuation, we obtain from (\ref{eq-H00}) the formula
\begin{equation} \label{eq-explicit}
G_\mu(z) = A_m \int_0^\infty  e^{-t(\mu-m)} \frac{t}{t^2 + (z+itv)^2} K_2 \big ( m \sqrt{t^2 + (z+ itv)^2} \big ) \diff t .
\end{equation} 
Here $K_2(z)$ stands for the modified Bessel function of the third kind, and $A_m >0$ denotes some constant. Notice that 
\begin{equation}
w = t^2 + (z + itv)^2 = (1-v^2) t^2 + z^2 +2 itv \cdot z
\end{equation} 
is a complex number with $| \mathrm{arg} \, w | < \pi/2$.

Next we analyze $G_\mu(z)$ for $|z| \leq 1$ and for $|z| > 1$ separately. From \cite{Abramowitz+Stegun1972} we recall the estimate
\begin{equation}
|K_2(mw)| \leq \frac{C}{|w|^2}, \quad \mbox{for $| \mathrm{arg} \, w | < \pi/2$},
\end{equation}  
which implies that $G_\mu(z)$ with $|z| \leq 1$ satisfies the bound
\begin{align*}
|G_\mu(z)| & \leq C \int_0^\infty e^{-t(\mu-m)} \Big | \frac{t}{(1-v^2) t^2 + |z|^2 + 2itv\cdot z} K_2 \big (m \sqrt{(1-v^2) t^2 + |z|^2 + 2itv \cdot z} \big ) \Big | \diff t \\
& \leq C \int_0^\infty e^{-t(\mu-m)} \frac{t}{[(1-v^2)t^2 + |z|^2]^2}  \diff t .
\end{align*}
Since $\mu - m \geq 0$, the $t$-integral is finite for $z \neq 0$ and we obtain
\begin{equation} \label{ineq-Gmu1}
|G_\mu(z)| \leq \frac{C}{|z|^2}, \quad \mbox{for $|z| \leq 1$},
\end{equation}
where we use that $|a+ib| \geq |a|$ and $|\sqrt{a + ib}| \geq \sqrt{|a|}$ holds for $a,b \in \RR$. 

To estimate $G_\mu(z)$ for $|z| > 1$, we use the bound
 \begin{equation}
|K_2(mw)| \leq C \Big | \frac{e^{-mw}}{|w|^2} \Big | \leq C \frac{e^{-m| \mathrm{Re} \, w |}}{|w|^2}, \quad \mbox{for $|\mathrm{arg} \, w| < \pi/2$ and $|w| > 1$} ,
\end{equation}  
taken from \cite{Abramowitz+Stegun1972}. By means of the inequality $\sqrt{a^2 + b^2} \geq (1-\epsilon) |a| + \epsilon |b|$, for any $ 0 < \epsilon \leq 1$, we proceed to find that
\begin{equation}
|G_\mu(z)|  \leq C e^{- \epsilon m |z|}  \int_0^\infty e^{-t (\mu-m+(1-\epsilon)\sqrt{1-v^2} m)} \frac{t}{[(1-v^2) t^2 + |z|^2]^2} \diff t , \quad \mbox{for $|z| \geq 1$} .
\end{equation} 
Our assumption on $\mu$ allows us to choose $\epsilon \in (0,1]$ such that exponent in the $t$-integral is nonpositive. The best $\epsilon$ is given by
\begin{equation} \label{eq-eps}
\epsilon = \min \left \{ 1, \frac{\Sigma_v + \mu}{m \sqrt{1-v^2}} \right \} \in (0, 1]
\end{equation}
and hence
\begin{equation} \label{ineq-Gmu2}
|G_\mu(z)| \leq C e^{- \epsilon m |z|} \int_0^\infty \frac{t}{[(1-v^2) t^2 + |z|^2]^2} \diff t \leq C \frac{e^{-\epsilon m |z|}}{|z|^2}, \quad \mbox{for $|z| \geq 1$}.
\end{equation} 
Combining now (\ref{ineq-Gmu1}) and (\ref{ineq-Gmu2}), we see that
\begin{equation} \label{ineq-Gmu}
|G_\mu(z)| \leq C \frac{e^{-m \epsilon |z|}}{|z|^2}, \quad \mbox{for $z \in \RR^3$,} 
\end{equation}
where $\epsilon$ is given by (\ref{eq-eps}) and $C$ is some constant. This shows that $G_\mu(z)$ exhibits exponential decay; in particular, we have that $G_\mu \in \Lp{p}(\RR^3)$ if $1 \leq p < 3/2$.

Returning to (\ref{eq-exp2}), we notice that 
\begin{equation} \label{eq-exp3}
\sol(x) = - \int_{\RR^3} G_\mu(x-y) V(y) \sol(y) \diff y .
\end{equation}
Moreover, the function $V(x) = -(|x|^{-1} \ast |\sol|^2)(x)$ obeys
\begin{equation} \label{eq-Vvanish}
V \in \C{0}(\RR^3) \quad \mbox{and} \quad \lim_{|x| \rightarrow \infty} V(x) = 0,
\end{equation}
since $f \ast g$ is a continuous function vanishing at infinity, provided that $f \in \Lp{p}$ and $g \in \Lp{p'}$ with $1/p + 1/p' = 1$ and $p>1$; see, \eg \cite{Lieb+LossII}. Here we note that, \eg $|x|^{-1} \in \Lp{2}(\RR^3) + \Lp{4}(\RR^3)$ and in particular $|\sol|^2 \in \Lp{4/3}(\RR^3) \cap \Lp{2}(\RR^3)$ since $\sol \in \Sob{s}(\RR^3)$ for all $s \geq 1/2$ (cf.~beginning of App.~C). 

Using (\ref{eq-exp3}), (\ref{ineq-Gmu}) and (\ref{eq-Vvanish}), pointwise exponential decay of $\sol(x)$ follows from a direct adaption of an argument by Slaggie and Wichmann for exponential decay of eigenfunctions for Schr\"odinger operators; see, \eg \cite{Hislop2000} for a convenient exposition of this method. This completes the proof of Lemma \ref{lem-exp-decay}.\end{proof}

\subsection*{Acknowledgments}
The authors are grateful to I.~M.~Sigal and M.~Struwe for useful discussions. E.~L.~also thanks D.~Christodoulou. 

\bibliographystyle{habbrv}
\bibliography{bosonbib}

\begin{thebibliography}{10}

\bibitem{Abramowitz+Stegun1972}
M.~Abramowitz and I.~A. Stegun, editors.
\newblock {\em Handbook of mathematical functions with formulas, graphs, and
  mathematical tables}.
\newblock Dover Publications Inc., New York, 1992.
\newblock Reprint of the 1972 edition.

\bibitem{Cazenave+Lions1982}
T.~Cazenave and P.-L. Lions.
\newblock Orbital stability of standing waves for some nonlinear
  {Schr\"odinger} equations.
\newblock {\em Comm. Math. Phys.}, 85:549--561, 1982.

\bibitem{Elgart+Schlein2005}
A.~Elgart and B.~Schlein.
\newblock Mean field dynamics of {Boson Stars}.
\newblock arXiv:math-ph/0504051, to appear in Comm.~Pure Appl.~Math., 2005.

\bibitem{FJL2005II}
J.~Fr{\"o}hlich, B.~L.~G. Jonsson, and E.~Lenzmann.
\newblock {Boson Stars: Effective dynamics in external potentials}.
\newblock In preparation, 2005.

\bibitem{Froehlich+Lenzmann2005}
J.~Fr{\"o}hlich and E.~Lenzmann.
\newblock Blow-up for nonlinear wave equations describing {Boson Stars}.
\newblock arXiv:math-ph/0511003, submitted, 2005.

\bibitem{Hislop2000}
P.~D. Hislop.
\newblock Exponential decay of two-body eigenfunctions: {A} review.
\newblock In {\em Proceedings of the Symposium on Mathematical Physics and
  Quantum Field Theory (Berkeley, CA, 1999)}, volume~4 of {\em Electron. J.
  Differ. Equ. Conf.}, pages 265--288 (electronic), San Marcos, TX, 2000.
  Southwest Texas State Univ.

\bibitem{Lenzmann2005LWP}
E.~Lenzmann.
\newblock Well-posedness for semi-relativistic {Hartree} equations of critical
  type.
\newblock arXiv: math-AP/0505456, 2005.

\bibitem{Lieb1977}
E.~H. Lieb.
\newblock Existence and uniqueness of the minimizing solution of {Choquard's}
  nonlinear equation.
\newblock {\em Stud. Appl. Math.}, 57:93--105, 1977.

\bibitem{Lieb+LossII}
E.~H. Lieb and M.~Loss.
\newblock {\em Analysis}, volume~14 of {\em Graduate Studies in Mathematics}.
\newblock AMS, second edition, 2001.

\bibitem{Lieb+Thirring1984}
E.~H. Lieb and W.~Thirring.
\newblock Gravitational collapse in quantum mechanics with relativistic kinetic
  energy.
\newblock {\em Ann. Physics}, 155(2):494--512, 1984.

\bibitem{Lieb+Yau1987}
E.~H. Lieb and H.-T. Yau.
\newblock The {Chandrasekhar} theory of stellar collapse as the limit of
  quantum mechanics.
\newblock {\em Comm. Math. Phys.}, 112:147--174, 1987.

\bibitem{Lions1984a}
P.-L. Lions.
\newblock The concentration-compactness principle in the calculus of
  variations. {T}he locally compact case, part {I}.
\newblock {\em Ann. Inst. Henri Poincar{\'e}}, 1(2):109--145, 1984.

\bibitem{Stein1993}
E.~M. Stein.
\newblock {\em Harmonic Analysis}.
\newblock Princeton University Press, Princeton, New Jersey, 1993.

\end{thebibliography}

\bigskip
\noindent 
{\sc J\"urg Fr\"ohlich \\ Institute for Theoretical Physics \\ ETH Zurich \\ 8093 Zurich, Switzerland.}\\
{\em E-mail address:} {\tt juerg@itp.phys.ethz.ch}

\bigskip
\noindent 
{\sc B.~Lars G.~Jonsson \\ Institute for Theoretical Physics \\ ETH Zurich \\ 8093 Zurich, Switzerland.} \vspace{1ex} \\
Alternative address: \\ {\sc Division of Electromagnetic Theory \\ Alfv\'en Laboratory \\ Royal Insitute of Technology \\ SE-100 44 Stockholm, Sweden.}\\
{\em E-mail address:} {\tt jonsson@itp.phys.ethz.ch}

\bigskip
\noindent
{\sc Enno Lenzmann \\ Department of Mathematics, HG G 33.1 \\ ETH Zurich \\ 8092 Zurich, Switzerland.}\\
{\em E-mail address:} {\tt lenzmann@math.ethz.ch}

\end{document}